\documentclass[twocolumn,aps,prl,superscriptaddress]{revtex4-2}
\usepackage{tabularx}
\usepackage{amsmath}
\usepackage[final]{hyperref}
\usepackage{ulem}
\hypersetup{
	colorlinks=true,
	linkcolor=blue,
	citecolor=blue,
	filecolor=magenta,
	urlcolor=blue
}
\usepackage{graphicx}
\usepackage{amssymb}
\usepackage{mathtools}
\usepackage{float,bm}
\usepackage{braket}

\begin{document}

\title{Spectral Topology and Non-Bloch Band Theory for Domain-Wall Systems}

\author{Mingtao Xu}
\affiliation{Laboratory of Quantum Information, University of Science and Technology of China, Hefei 230026, China}
\author{Rui Wang}
\affiliation{School of Physics and Astronomy, Yunnan Key Laboratory for Quantum Information,
Yunnan University, Kunming 650091, China}
\author{Tian-Shu Deng}
\email{20220225@ynu.edu.cn}
\affiliation{School of Physics and Astronomy, Yunnan Key Laboratory for Quantum Information,
Yunnan University, Kunming 650091, China}
\author{Wei Yi}
\email{wyiz@ustc.edu.cn}
\affiliation{Laboratory of Quantum Information, University of Science and Technology of China, Hefei 230026, China}
\affiliation{Anhui Province Key Laboratory of Quantum Network, University of Science and Technology of China, Hefei 230026, China}
\affiliation{CAS Center For Excellence in Quantum Information and Quantum Physics, Hefei 230026, China}
\affiliation{Hefei National Laboratory, University of Science and Technology of China, Hefei 230088, China}
\date{\today}

\begin{abstract}
We study the spectral topology of one-dimensional non-Hermitian models in a domain-wall configuration, where different domains are arranged in a ring geometry. 
While eigenstates can localize near an interface under the non-Hermitian skin effect, we show that the localization of an eigenstate originates from the difference in the spectral winding numbers,  with respect to the corresponding eigenenergy, between the two adjacent domains.
We then obtain the conditions for the generalized Brillouin zone (GBZ) in the complex momentum space, 
by extending the Ronkin-function formalism to the domain-wall configuration.
In addition to the conventional skin modes that correspond to standing waves on individual domains under the open boundary condition, a unique type of traveling-wave-like skin modes emerges, whose construction involves all domains. Besides their difference in the spatial profiles, these two types of modes obey distinct GBZ conditions, making them differentiable on the GBZ. Interestingly, the traveling-wave-like modes further carry a finite flux spectral winding number, indicating their boundary sensitivity. 
\end{abstract}

\maketitle

{\it Introduction.---}
Boundary conditions play an important role in non-Hermitian systems. A paradigmatic example is the non-Hermitian skin effect (NHSE)~\cite{YaoEdge,HatanoLocalization}, where an extensive number of eigenstates become localized near boundaries, and the open-boundary spectrum differs drastically from that under the periodic boundary condition (PBC).  
For a translationally invariant system, this phenomenon is well-understood through the spectral topology~\cite{KawabataSymmetry,ZhangCorrespondence,OkumaTopological,HuTopological} and the non-Bloch band theory~\cite{KunstBiorthogonal,GongTopological,AshidaNon,OkumaNonreview,YokomizoNon,SongNon,Yokomizotwo,YangNon,KunstNon,YaoChern,WangAmoeba,XiongNon,HuGeometric,ZhangAlgebraic}. The spectral winding under the PBC anticipates NHSE under the open-boundary condition (OBC), while the generalized Brillouin zone (GBZ) determines the spatial localization and eigenspectrum under the OBC.

But boundary conditions are not limited to the PBC and OBC. The domain-wall (DW) boundary, where two bulks (or, equivalently, two domains) are brought to contact through an interface, is ubiquitous and naturally accessible to quantum simulation platforms~\cite{XiaoNP,XiaoNon,XiaoObservation,LinObservation,LiuExperimental,ZhangObservation,XiaoObservation2,XiaoObservation3}. It is in fact a generalization of both the OBC and PBC over a single bulk. Replacing a bulk with the trivial vacuum in a DW configuration would lead to the OBC, whereas the PBC holds for a DW configuration in a ring geometry [see Fig.~\ref{fig1}(a)], despite the lack of lattice translational invariance.
Nevertheless, studies of the spectral topology and non-Bloch band theory of DW configurations are so far limited to the simplest scenario of two domains~\cite{EzawaNonH,DengNon,DengNon2,LiBand}, and the description therein remains phenomenological.

In this work,  we study the spectral topology and non-Bloch description of 
a general one-dimensional non-Hermitian system composed of $n$ domains in a ring geometry [dubbed the $n$-domain-wall ($n$-DW) ring, sketched in Fig.~\ref{fig1}(a)]. 
We first establish that the NHSE in a DW ring is signaled not by the spectral winding of any isolated bulk, but by the winding-number mismatch between adjacent domains. 
The presence of skin modes at a given interface is protected by a nonzero relative winding. 
Further analysis through the Ronkin functions in the complex momentum space~\cite{HuTopological,WangAmoeba,XiongNon,MikhalkinAmoeba,YangObservation,XuOptimal,ZhongUnveiling} reveals that the DW spectrum is generally divided into the standing-wave- and traveling-wave-like sectors [see Fig.~\ref{fig1}(b)]. 
The standing-wave-like sector features a spectral overlap with those constructed from the GBZs or auxiliary-GBZs (aGBZ) of individual domains~\cite{YangNon}, and physically corresponds to standing waves formed over a given domain (though with complex momenta and hence spatial decay or amplification in amplitude).
By contrast, the traveling-wave-like sector has no spectral overlap with any individual domains under the OBC, but satisfies a global PBC as the ring is traversed.
As such, the eigenstates over a DW ring can be categorized into three general types: i) those unaffected by the NHSE, as schematically illustrated in Fig.~\ref{fig1}(c); ii) the standing-wave-like skin modes, with exponential decay enveloping spatial amplitude oscillations [see Fig.~\ref{fig1}(d)]; and iii) the traveling-wave-like skin modes, with exponential amplitude decay, as illustrated in  Fig.~\ref{fig1}(e).
Note that while type i) and ii) are also found for non-Hermitian models under the OBC, type iii) is unique to the DW rings. 
We show that a complete characterization of all eigenstates can thus be achieved combining the relative windings and Ronkin-function calculations. 
Furthermore, the traveling-wave-like modes generally carry a finite flux spectral winding number, indicating their boundary sensitivity. 
Our work fully resolves the topological origin of non-Hermitian skin effects in domain-wall systems, and  establishes the corresponding non-Bloch band theory.

\begin{figure}
    \centering
    \includegraphics[width=1\linewidth]{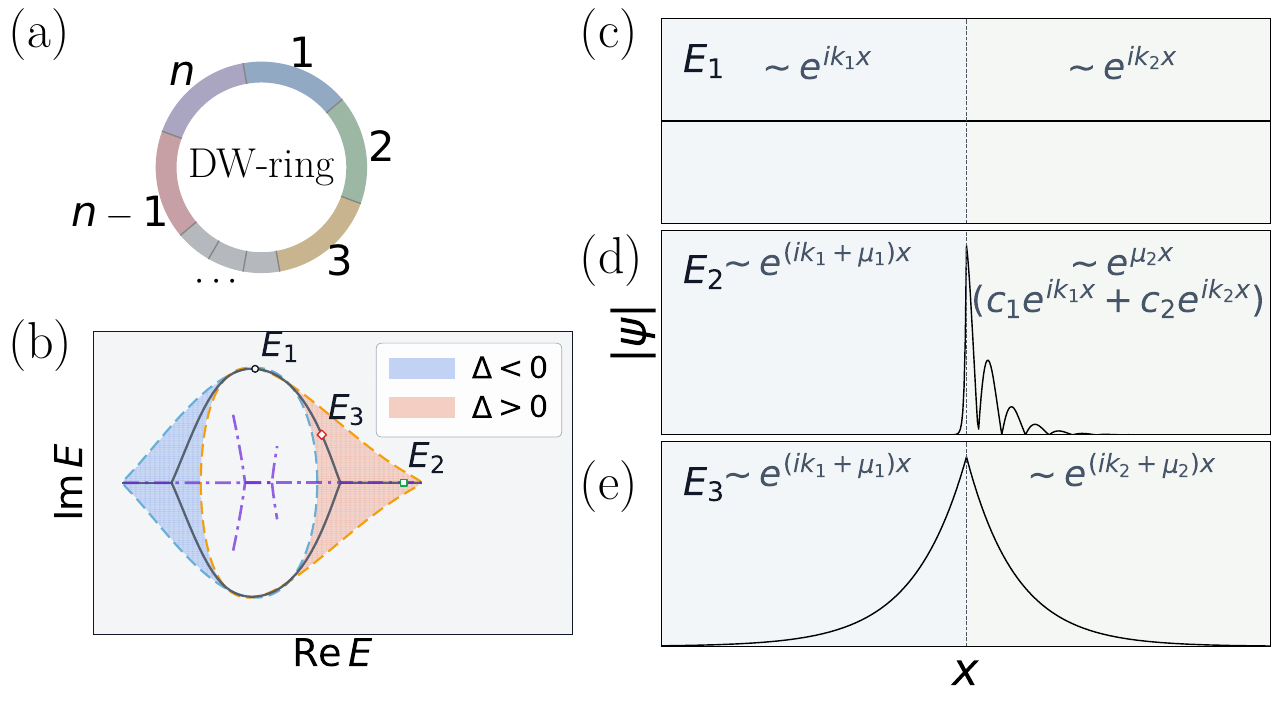}
    \caption{Overview of the DW non-Bloch theory. (a) Schematic of an $n$-DW ring. (b) Schematic eigenspectrum of a $2$-domain ring. The dashed curves denote the PBC spectra of the two constituent domains, the solid black curves denote the DW-ring spectrum, and the dash-dotted purple curves denote the spectrum obtained from the individual-domain (a)GBZ conditions. The color-shaded regions (bounded by dashed curves) indicate different signs of the winding-number mismatch $\Delta(E)$, whereas $\Delta(E)=0$ in other regions.
(c)(d)(e) Typical spatial wave-function profiles for (c) extended state without NHSE, (d) standing-wave-like skin modes, and (e) traveling-wave-like skin modes, with their eigenenergies marked as $E_1$, $E_2$, and $E_3$ in (b), respectively.} 
    \label{fig1}
\end{figure}

{\it Topological origin of NHSE.---}
As the foundation of our analysis, we first consider a single interface between the $\alpha$th and $(\alpha+1)$th domain, denoted by $\alpha|\alpha+1$. The spectral winding of domain $\alpha$ with respect to an arbitrary complex energy $E$ is given by~\cite{GongTopological,KawabataSymmetry}
\begin{equation}
    w_\alpha(E)
    =
    \frac{1}{2\pi i}
    \int_0^{2\pi} dk\,
    \partial_k
    \log\det[h_\alpha(e^{ik})-E],\label{eq:winding}
\end{equation}
where $h_\alpha(e^{ik})$ is the Bloch Hamiltonian of the domain, with $k\in [0,2\pi)$.
The spectral winding is well defined as long as the point gap at $E$ remains open, namely, when $\det[h_\alpha(e^{ik})-E]\neq 0$ for all $k\in[0,2\pi)$. We then define the relative spectral winding at the interface $\alpha|\alpha+1$ as
\begin{equation}
    \Delta_\alpha(E)
    =
    w_{\alpha+1}(E)-w_\alpha(E).\label{eq:diffw}
\end{equation}

Our first key result is that NHSE in a DW ring is topologically protected by the winding mismatch between adjacent domains. In particular, an eigenstate with energy $E$ localized near $\alpha|\alpha+1$ necessarily requires $\Delta_\alpha(E)>0$, provided that both adjacent winding numbers are well defined.
Therefore, a DW-ring eigenstate can exhibit NHSE only if a positive relative winding exists. Equivalently, if no such mismatch exists, the eigenstate has no NHSE.
This result represents the topological origin of the NHSE in a DW configuration, which is reminiscent of the case for a single bulk with OBC~\cite{ZhangCorrespondence,OkumaTopological}, and can be proved similarly~\cite{SM}. Indeed, in the special case that one domain becomes a trivial vacuum, the interface is replaced by an open boundary, and the relative winding reduces to the bulk spectral winding, recovering the conventional OBC case.

\begin{figure*}
    \centering
    \includegraphics[width=1\linewidth]{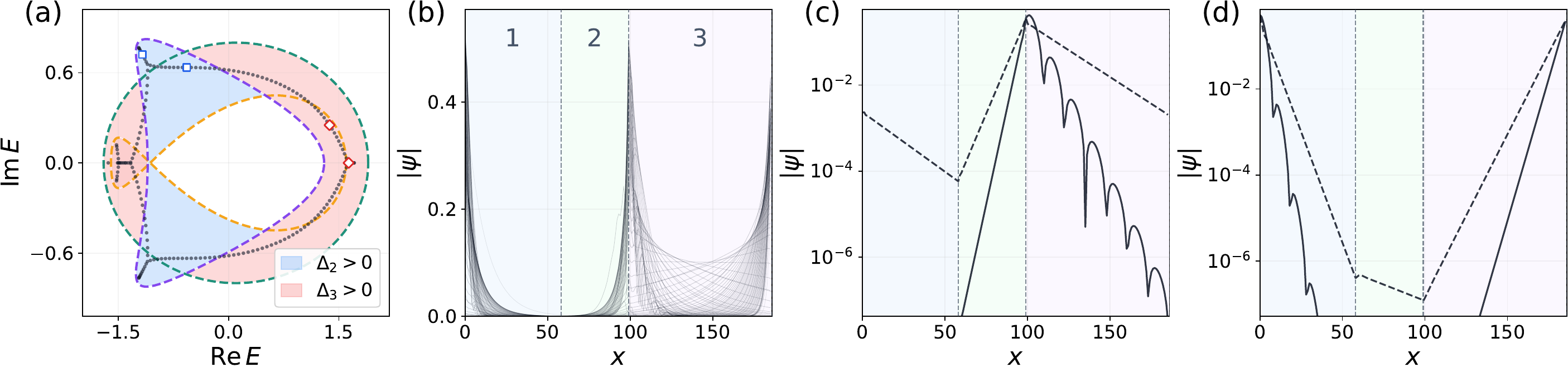}
    \caption{
Topological origin of NHSE in a $3$-DW model.
(a) PBC spectra of the three constituent domains and eigenenergies of the full DW-ring. The green, orange, and purple dashed curves denote the PBC spectra of domains 1, 2, and 3, respectively, while the black dots denote the DW-ring eigenenergies. The shaded regions indicate $\Delta_2>0$ (blue) and $\Delta_3>0$ (red), for the interfaces $2|3$ and $3|1$, respectively, while we have $\Delta_2=\Delta_3=0$ in regions not shaded. The blue markers correspond to the parameters in (c), whereas the red markers correspond to those in (d).
(b) Spatial wave-function profiles for all eigenstates.
(c)(d) Spatial profiles of representative right eigenstates localized near the $2|3$ and $3|1$
interfaces. The solid and dashed curves correspond to the standing-wave- and
traveling-wave-like sectors, respectively. The model is given by
$h_1(\beta)=0.5\beta^{-1}+0.1+1.3\beta$,
$h_2(\beta)=0.7\beta^{-1}-0.3+0.9\beta+0.3\beta^2$, and
$h_3(\beta)=0.3\beta^{-2}+0.2\beta^{-1}-0.4+1.0\beta+0.2\beta^2$,
with $N_1:N_2:N_3=58:41:87$.
}
    \label{fig2}
\end{figure*}

As an example, we consider a $3$-DW ring, where the bulk Hamiltonians are
\begin{align}
    h_1(\beta) &=
    t_{1,-1}\beta^{-1}+t_{1,0}+t_{1,1}\beta,\\
    h_2(\beta)
    &=
    t_{2,-1}\beta^{-1}+t_{2,0}+t_{2,1}\beta+t_{2,2}\beta^2,\\
    h_3(\beta)
    &=t_{3,-2}\beta^{-2}+t_{3,-1}\beta^{-1}+t_{3,0}+t_{3,1}\beta+t_{3,2}\beta^2.
\end{align}
The full Hamiltonian $H_{\rm DW}$ is obtained by connecting these domains through local hoppings across the interfaces. As shown in Fig.~\ref{fig2}, all eigenstates of the system are localized at one of the interfaces:
those localized at $2|3$ have eigenenergies in the blue region of Fig.~\ref{fig2}(a), where $\Delta_2>0$, whereas those localized at $3|1$ have eigenenergies in the red region, where $\Delta_3>0$.
Further examination of the spatial profiles of the wave functions indicate that two types of skin modes exist. Those featuring a simple exponential decay in each domain [Fig.~\ref{fig2}(c)(d) dashed], corresponding to the traveling-wave-like skin modes in Fig.~\ref{fig1}(e). And those 
with finer oscillations [Fig.~\ref{fig2}(c)(d) solid], corresponding to the standing-wave-like skin modes in Fig.~\ref{fig1}(d).
Interestingly, while traveling-wave-type eigenstates are absent in the simplest single-bulk OBC case, a large number of them generally emerge in a DW configuration.

Here two questions arise. How to differentiate these two types of skin modes? And, more fundamentally, how to determine whether the complex energy $E$ lies within the eigenspectrum of the DW ring? 
As we show below, these can be addressed by devising Ronkin functions for the DW-ring configuration, which provide a shortcut to the non-Bloch band theory for DW rings.

{\it Non-Bloch band theory through the Ronkin function.---}
For an eigenstate on the DW ring in the thermodynamic limit, regardless of its type, there necessarily exists a dominant decay rate for its spatial wave-function profile within each domain. Denoting the decay rate of the $\alpha$th domain as $\kappa_\alpha$, the NHSE is present if $\exists \alpha$, $\kappa_\alpha\neq 0$. Calculating $\{\kappa_\alpha\}$ for all eigenstates enables the construction of the GBZ, as well as the eigenspectrum of the system in the thermodynamic limit, both being the central tasks of a non-Bloch band theory. 

For a qualitative understanding of the process, we consider a domain-dependent imaginary gauge transformation $V_{\boldsymbol\mu}$ such that
\begin{equation}
    |V_{\boldsymbol\mu}^{-1}\psi_\alpha(x)|
    =
    e^{-\mu_\alpha x}|\psi_\alpha(x)|
    \sim e^{(\kappa_\alpha-\mu_\alpha)x},
\end{equation}
where $\psi_\alpha(x)$ represents the corresponding eigen wave function. 
Naturally, if the transformed state is no longer a skin mode, we should have $\mu_\alpha=\kappa_\alpha$, so that the decay in domain $\alpha$ is compensated by $\mu_\alpha$. Considering the single-valuedness of the transformed wave function, we impose the global constraint $\sum_\alpha r_\alpha\mu_\alpha=0$, where $\boldsymbol\mu=(\mu_1,\ldots,\mu_{n-1})$ with $\mu_n=-\sum_{\alpha=1}^{n-1}(r_\alpha/r_n)\mu_\alpha$~\cite{SM}. Here $ r_\alpha= N_\alpha/\sum_\gamma N_\gamma$ and $N_\alpha$ is the number of unit cells of domain $\alpha$.

According to the topological criterion established above, given a complex reference energy $E$, the absence of skin modes thereof in the transformed model requires
\begin{equation}
    w_1(\mu_1;E)=w_2(\mu_2;E)=\cdots=w_n(\mu_n;E),
    \label{eq:windeq}
\end{equation}
where the point gaps and winding numbers are defined with the transformed Hamiltonians $h_\alpha(e^{ik_\alpha+\mu_\alpha})$. Note that $k_\alpha-i\mu_\alpha$ parameterizes the complex momentum space in the non-Bloch band theory for domain $\alpha$~\cite{YaoEdge,YokomizoNon}.

One then expects to solve $\mu_\alpha$ and $E$ from Eq.~(\ref{eq:windeq}) under the global constraint, for the decay rates $\kappa_\alpha$ and the corresponding eigenenergies of the DW configuration.
However, an infinite number of solutions generally exist, though not all of them satisfy the eigen problem of the DW ring.
When $E$ lies outside the DW-ring eigenspectrum, Eq.~(\ref{eq:windeq}) admits a continuous set of solutions $\boldsymbol\mu$, corresponding to different imaginary gauges that remove all winding mismatches. As $\boldsymbol\mu$ continuously varies across this solution space, the point gap of the transformed non-Bloch Hamiltonian remains open with respect to $E$, except at the boundary of the solution space where the point gap closes.
As $E$ asymptotically approaches the DW-ring eigenspectrum, the interior of the solution space collapses, and, in at least one of the domains ($\alpha$ for instance), the corresponding $\mu_\alpha$ approaches the physical decay rate $\kappa_\alpha$, as the point gap of the transformed Hamiltonian asymptotically closes. 
Inspired by the single-domain Ronkin-function formalism under the OBC~\cite{WangAmoeba}, we characterize this critical limit by the constrained Ronkin function.

\begin{figure*}
    \centering
    \includegraphics[width=1\linewidth]{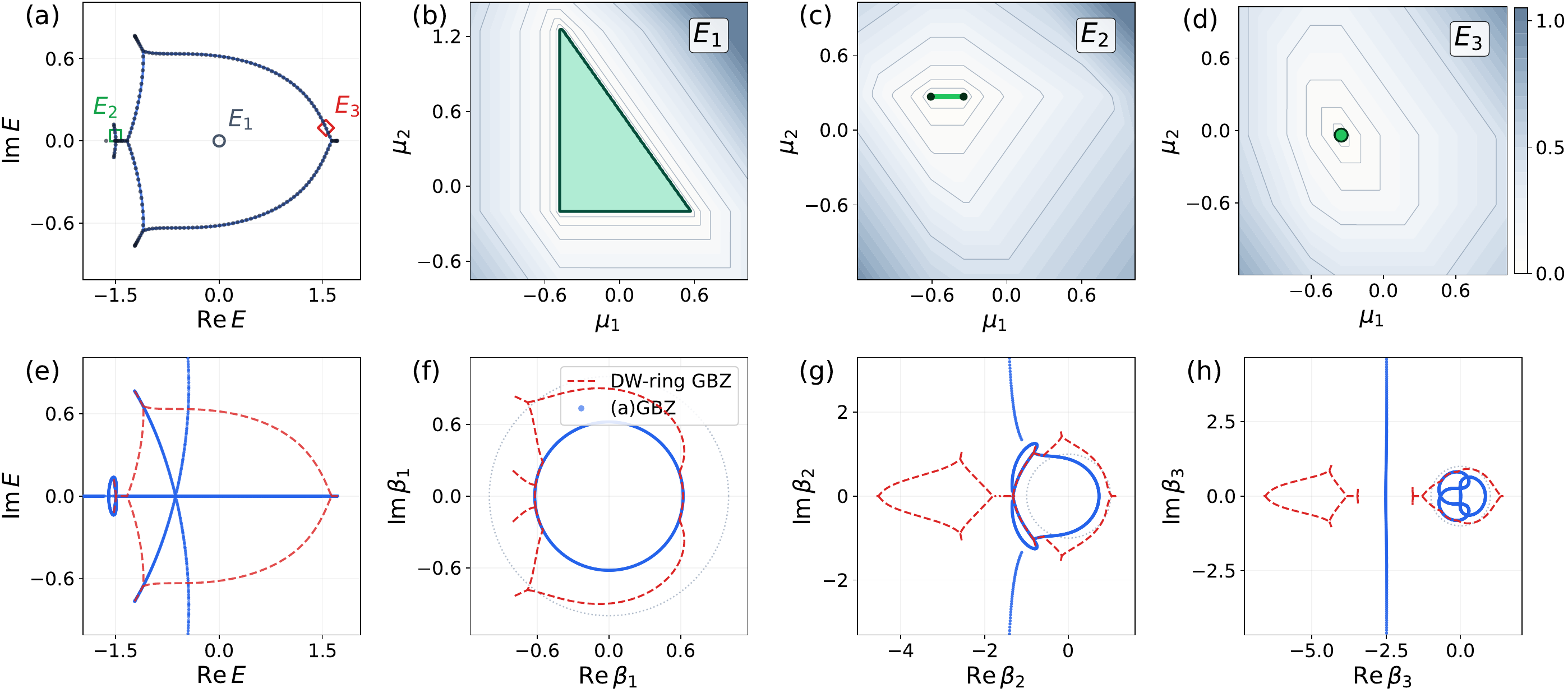}
    \caption{Ronkin function and domain-resolved GBZs for the $3$-DW ring. (a) DW-ring eigenspectrum obtained from direct diagonalization (dotted) and the Ronkin-function calculation (solid). 
    (b)(c)(d) Constrained Ronkin functions at $E_1$, $E_2$, and $E_3$, as marked in (a), respectively. 
  Typically, a two-dimensional flat region of the Ronkin function emerges, as in (b), indicating that 
  $E_1$ lies in a point gap of the DW-ring. 
 The collapse of the flat region, illustrated in (c) and (d), places the corresponding $E_2$ and $E_3$ on the DW eigenspectrum. Here, (c) corresponds to a standing-wave-like skin mode and (d) corresponds 
 to a traveling-wave-like skin mode.  
 (e) Comparison between the DW-ring spectrum (red) and the individual-domain (a)GBZ spectra (blue). 
 Eigenenergies of the standing-wave-like modes overlap with those of the individual-domain (a)GBZ spectra.
 (f)(g)(h) Domain-resolved GBZs in the complex $\beta_\alpha$ plane. The DW-ring GBZ is generally multi-valued and differs from the isolated-domain (a)GBZ. The model and parameters are the same as those in Fig.~\ref{fig2}.}
    \label{fig3}
\end{figure*}

We define the constrained Ronkin function for an $n$-DW ring
\begin{equation}
    \mathcal R(\boldsymbol\mu;E)
    =
    \sum_{\alpha=1}^{n-1}
    r_\alpha R_\alpha(\mu_\alpha;E)
    +
    r_n
    R_n\left(
    -\sum_{\alpha=1}^{n-1}
    \frac{r_\alpha}{r_n}\mu_\alpha;E
    \right),
    \label{eq:ronkin}
\end{equation}
where the Ronkin function in domain $\alpha$ is
\begin{equation}
    R_\alpha(\mu_\alpha;E)
    =
    \int_0^{2\pi}
    \frac{dk_\alpha}{2\pi}
    \log
    \left|
 f_\alpha(\beta_\alpha;E)
    \right|,
    \label{eq:ronkin_a}
\end{equation}
and $\beta_\alpha:=e^{ik_\alpha+\mu_\alpha}$.
For each domain, $f_\alpha(\beta_\alpha;E)=\det[h_\alpha(\beta_\alpha)-E]$ gives a Laurent series in $\beta_\alpha$, with $f_\alpha(\beta_\alpha;E)=\sum_{n=-s_\alpha}^{p_\alpha}a_{\alpha,n}(E) \beta_\alpha^n$, where $s_\alpha$ represents the order of the pole at $\beta_\alpha=0$.
Sorting the roots $\beta_{\alpha,m}(E)$ of the characteristic equation $f_\alpha(\beta_\alpha;E)=0$ by their modulus, we have
\begin{align}
\mu_{\alpha,1}(E)\le \mu_{\alpha,2}(E)\le \cdots\le\mu_{\alpha,s_\alpha+p_\alpha}(E),
\end{align}
where $\mu_{\alpha,m}=\log|\beta_{\alpha,m}|$. 
By construction, $R_\alpha(\mu_\alpha,E)$ is a piecewise linear function, and its derivative with respect to $\mu_\alpha$ equals the spectral winding of the corresponding non-Bloch Hamiltonian~\cite{WangAmoeba}. Hence, we have $\partial_{\mu_\alpha}R_\alpha(\mu_\alpha;E)= w_\alpha(\mu_\alpha;E)$ for
 $\mu_\alpha\in \left(\mu_{\alpha,m_\alpha(w_\alpha)},\mu_{\alpha,m_\alpha(w_\alpha)+1}\right)$, where the root index $m_\alpha(w_\alpha)=s_\alpha+w_\alpha$. Note that we take $\mu_{\alpha,0}=-\infty$, $\mu_{\alpha,s_\alpha+p_\alpha+1}=+\infty$ for convenience.

Importantly, the derivative of the constrained Ronkin function Eq.~(\ref{eq:ronkin}) with respect to $\mu_\alpha$ is ($\alpha\neq n$)
\begin{align}
    \partial_{\mu_\alpha}\mathcal R
    =
    r_\alpha
    \left[
    w_\alpha(\mu_\alpha;E)
    -
    w_n(\mu_n;E)
    \right],
\end{align}
so that the solutions to Eq.~(\ref{eq:windeq}) corresponds to a flat region in $\mathcal{R}(\boldsymbol\mu,E)$, characterized by $\partial_{\mu_\alpha}\mathcal{R}=0$, within an $(n-1)$-dimensional $\boldsymbol \mu$ space (that is, the solution space).
At the boundary of this flat region, the point gap of the transformed non-Bloch Hamiltonian closes for one or more domains, and $E$ belongs to the corresponding transformed PBC spectra. However, $E$ only lies on the DW-ring eigenspectrum when the interior of the flat region collapses, which corresponds to the aforementioned critical limit. Identification of this critical scenario enables the solution of the DW eigenenergies $E$, which are prerequisite for solving $\kappa_\alpha$ and constructing the GBZ. The result is summarized as the following two cases~\cite{SM}, which are essentially the DW-ring GBZ condition:
\begin{align}
  &  \text{Case \rm I,} \,\,\quad \exists w,\,\text{such that}\notag \\
    &\quad \quad \qquad\begin{cases}
        \sum_{\alpha}r_\alpha \mu_{\alpha,m_\alpha(w)}\le 0 \le \sum_{\alpha}r_\alpha \mu_{\alpha,m_\alpha(w)+1},\\
        \exists \alpha,\,\, \mu_{\alpha,m_\alpha(w)} = \mu_{\alpha,m_{\alpha}(w)+1},
    \end{cases}\nonumber\\
     &\text{Case \rm II,} \quad\exists w,\,\text{such that}\,\, \sum_{\alpha}r_\alpha \mu_{\alpha,m_\alpha(w)}=0 .\nonumber
\end{align}

Physically, Case I gives rise to the standing-wave-like modes. Here, at least one domain $\alpha$ satisfies the standing-wave condition $\mu_{\alpha,m_\alpha} = \mu_{\alpha,m_{\alpha}+1}$, which is the natural DW-ring generalization of the standard GBZ or aGBZ condition in a single bulk under the OBC~\cite{YaoEdge,YokomizoNon,YangNon}.
The additional condition $\sum_{\alpha}r_\alpha \mu_{\alpha,m_\alpha}\le 0 \le \sum_{\alpha}r_\alpha \mu_{\alpha,m_\alpha+1}$ ensures that this local standing-wave structure is compatible with the global constraint $\sum_\alpha r_\alpha\mu_{\alpha}=0$. 

By contrast, Case II yields the traveling-wave-like solution. In this case, each domain contributes one dominant non-Bloch mode with decay rate $\mu_{\alpha,m_\alpha}$. Although the wave function may grow or decay inside each individual domain, the PBC across the DW ring is ensured by $\sum_{\alpha}r_\alpha\mu_{\alpha,m_\alpha}=0$. Such a case is unique to the DW ring, and is a DW-ring generalization of the Bloch wave. 

From the perspective of the Ronkin function, while Case I merely reduces the $\boldsymbol \mu$-space dimension of the flat region, Case II pins all $\mu_\alpha$ to an isolated point. 
In either case, the GBZ conditions, combined with the characteristic equations $f_\alpha(\beta_\alpha;E)=0$, fully determine the solutions of the DW-ring eigenspectrum and its GBZ. In particular, for each eigenenergy $E$ on the DW spectrum, these equations give \(\beta_\alpha(E)=e^{ik_\alpha(E)+\kappa_\alpha(E)}\), enabling the construction of the domain-resolved GBZ.
As a result, each domain has its own GBZ, and the full DW eigenspectrum can be constructed from any single GBZ of the constituent domains.

With these understandings, the difference between the standing-wave-like and traveling-wave-like modes can be identified in the eigenspectrum and their spatial profiles. First, since the conditions for Case I are consistent with either the GBZ or aGBZ condition of the corresponding domains, the eigenspectrum of the standing-wave-like modes of a DW ring necessarily overlaps with the spectrum given by the GBZ or aGBZ condition of a constituent domain. The remaining non-overlapping spectrum
must belong to the traveling-wave-like
modes. Second, while both modes feature an exponential decaying (or growing) spatial profile, the standing-wave-like modes exhibit spatial modulations in amplitude enveloped by the exponential profile [see Fig.~\ref{fig1}(d)].
Note that the extended states without NHSE in Fig.~\ref{fig1}(c) satisfy the criteria for either Case I or Case II, although their spatial profile remains uniform due to $\mu_{\alpha,m_\alpha(w)}=0$.

Figure~\ref{fig3} demonstrates the constrained Ronkin construction for our $3$-DW-ring model. A finite flat region indicates that the imaginary gauge is not uniquely fixed, and the reference energy is outside the DW-ring eigenspectrum [Fig.~\ref{fig3}(b)]. 
It is only when the flat region collapses that the reference energy lies on the eigenspectrum [Fig.~\ref{fig3}(c)(d)].
By scanning the complex-energy plane with this criterion, we reconstruct the DW-ring spectrum; further substitution of the eigenenergies into the characteristic equations yields the domain-resolved $\rm GBZ$.
On each domain-resolved GBZ, standing-wave-like and traveling-wave-like modes can be further distinguished [see Fig.~\ref{fig3}(e)(f)(g)(h)]. Specifically, like the DW eigenspectrum, for the GBZ of each domain, sectors overlapping with the individual-domain (a)GBZ correspond to standing-wave-like modes, whereas the remaining sectors correspond to the traveling-wave-like modes. 

{\it Spectral winding and boundary sensitivity.---}
In the case where hoppings across one of the interfaces vanish, leading to a DW chain with OBC on its two ends, only the standing-wave-like eigenmodes survive, along with the conventional GBZ condition (Case I)~\cite{SM}. Hence the non-Hermitian boundary sensitivity enforces the transformation of traveling-wave-like modes to standing-wave-like modes. The phenomenon can be characterized by inserting a flux $\Phi$ through the ring, and defining the flux spectral winding number~\cite{GongTopological,ShnerbWinding,LiuLocalization,LonghiTopological,ClaesSkin,AcharyaLocalization,SunLyapunov}
\begin{equation}
    W_{\rm DW}(E_B)
    =
    \frac{1}{2\pi i}
    \int_0^{2\pi}d\Phi\,
    \partial_\Phi
    \log\det[H_{\rm DW}(\Phi)-E_B],
\end{equation}
where $E_B$ remains in a point gap for all $\Phi$, and $H_{\rm DW}(\Phi)$ is obtained by modulating each hopping amplitude from site $j$ to site $i$ with a distance-dependent phase $e^{i \Phi(i-j)/\sum_\gamma N_\gamma}$. This flux spectral winding measures the spectral flow of the closed DW ring under a $2\pi$ flux insertion. A nonzero $W_{\rm DW}$ implies that the spectrum winds around $E_B$, and cannot be continuously deformed to that of the corresponding open DW chain without closing the point gap.

Importantly, as illustrated in Fig.~\ref{fig4}, the nonzero $W_{\rm DW}$ region is bounded by the eigenspectrum of the traveling-wave-like modes. This is because a traveling-wave-like state is fixed by the phase accumulated after a full round trip across the ring, and the inserted flux directly shifts this phase~\cite{SM}. By contrast, standing-wave-like states are fixed by local equal-modulus conditions within individual domains, and do not contribute to the flux spectral winding in the thermodynamic limit. 
Such a spectral winding signals the boundary sensitivity of the traveling-wave-like modes.

\begin{figure}
    \centering
    \includegraphics[width=0.85\linewidth]{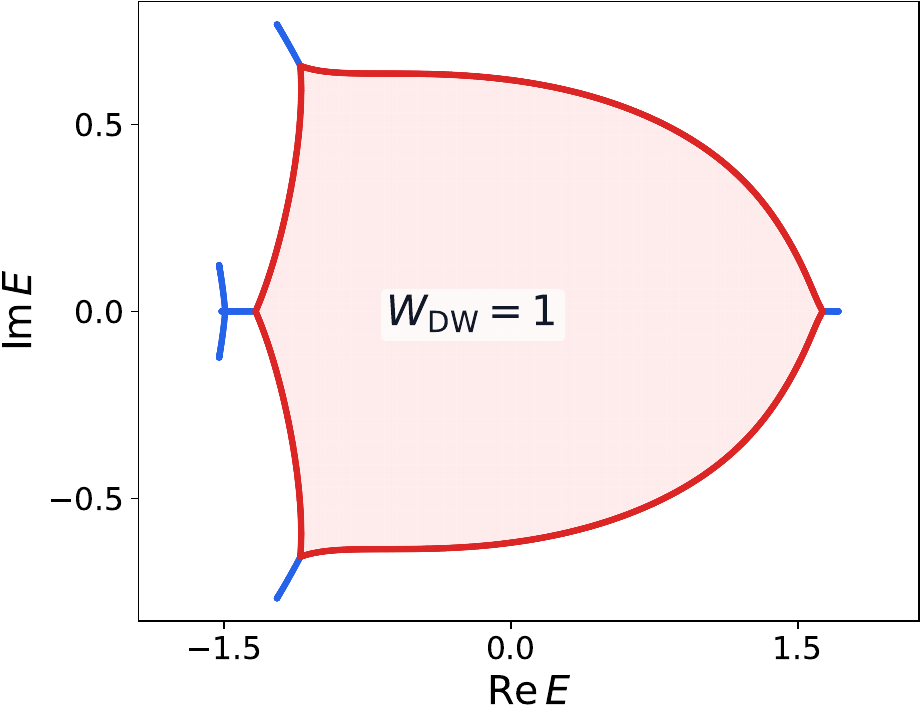}
    \caption{Spectral winding and eigenspectrum of a $3$-DW ring.
The shaded region features nonzero flux spectral winding
$W_{\rm DW}=1$.
The red curves represent eigenenergies of the traveling-wave-like modes, 
and the blue curves are those of the standing-wave-like modes.
The model and parameters are the same as those in Fig.~\ref{fig2}.}
    \label{fig4}
\end{figure}

{\it Discussion.---}
We have established the topological origin of NHSE in the DW-ring configuration, and recovered the non-Bloch band theory by constructing the GBZs. Unique to the DW-ring configuration, the majority of the skin modes therein belong to the traveling-wave type, satisfying the PBC in the DW-ring configuration and with distinct spatial profiles and GBZ conditions compared to the standing-wave-like skin modes that are commonly seen in a single bulk under the OBC. We discuss how these different skin modes can be differentiated by their features on the eigenspectrum and the GBZ, and demonstrate the presence of a flux spectral winding for the traveling-wave-like modes, which underlies the boundary sensitivity of these states.
By offering a thorough understanding of the skin modes and their non-Bloch descriptions in the DW configurations, our study paves the way for further investigations of the rich dynamic phenomena in DW systems, which are easily accessible in a variety of experimental platforms.

This work is supported by the National Natural Science Foundation of China (Grant Nos. 12374479 and 12304567), Quantum Science and Technology-National Science and Technology Major Project (Grant No. 2021ZD0301205), and by the Yunnan Fundamental Research Projects (Grant No. 202401CF070187).

\makeatletter
\let\MainBibSection\bibsection
\renewcommand{\bibsection}{%
  \let\SavedAddcontentsline\addcontentsline
  \let\addcontentsline\@gobblethree
  \MainBibSection
  \let\addcontentsline\SavedAddcontentsline
}
\makeatother

\makeatletter
\let\bibsection\MainBibSection
\makeatother

\widetext
\pagebreak

\renewcommand{\theequation}{S\arabic{equation}}
\renewcommand{\thefigure}{S\arabic{figure}}
\renewcommand{\thetable}{S\arabic{table}}
\setcounter{equation}{0}
\setcounter{figure}{0}
\setcounter{table}{0}

\begin{center}
{\bf \large Supplemental Material for\\
``Spectral Topology and Non-Bloch Band Theory for Domain-Wall Systems''}
\end{center}

\maketitle

\tableofcontents

\section{SI.~The Domain-Wall Ring Imaginary Gauge Transformation}

Here we clarify why the piecewise imaginary gauge used in the main text must satisfy the zero-growth condition on a closed domain-wall(DW) ring.  Let $x_\alpha$ be the coordinate inside domain $\alpha$, with length $N_\alpha$.  A piecewise imaginary gauge transformation acts on the wave function in each domain as
\begin{equation}
    \tilde{\psi}_\alpha(x_\alpha)
    =
    V_{\boldsymbol\mu}^{-1}(x_\alpha)\psi_\alpha(x_\alpha)
    =
    e^{-\mu_\alpha x_\alpha}\psi_\alpha(x_\alpha),
\end{equation}
where \(x_\alpha\) denotes the local coordinate in domain \(\alpha\). We then have $V_{\boldsymbol\mu}^{-1}(x_\alpha)=e^{-\mu_\alpha x_\alpha}$. For a ring geometry, the physical wave function satisfies the periodic boundary condition, which requires the gauge-transformed wave function \(\tilde{\psi}\) to obey the same periodic boundary condition. This condition gives
\begin{equation}
    V_{\boldsymbol{\mu}}^{-1}\bigg(x_\alpha+\sum_{\gamma=1}^n N_\gamma\bigg)
    =
    V_{\boldsymbol{\mu}}^{-1}(x_\alpha).
\end{equation}
On the other hand, because the gauge transformation is applied piecewise, we have
\begin{equation}
    V_{\boldsymbol{\mu}}^{-1}\bigg(x_\alpha+\sum_{\gamma=1}^n N_\gamma\bigg)
    =
    \exp \bigg(-\sum_{\gamma=1}^n N_\gamma\mu_\gamma\bigg)
    V_{\boldsymbol{\mu}}^{-1}(x_\alpha).
\end{equation}
Therefore, the piecewise imaginary gauge transformation is compatible with the periodic boundary condition only if the total imaginary flux satisfies
\begin{equation}
    \sum_{\gamma=1}^n N_\gamma\mu_\gamma =0 .
\end{equation}

For a thermodynamic eigenstate on the continuous band, each domain has a well-defined dominant exponential envelope characterized by a decay rate $\kappa_\alpha$. After one round trip around the closed ring, the dominant envelope accumulates a factor $\exp\left(\sum_{\alpha=1}^{n}N_\alpha\kappa_\alpha\right)$. The periodic boundary condition requires the exponential growth after one round trip to vanish, which gives
\begin{equation}
    \sum_{\gamma=1}^{n}N_\gamma\kappa_\gamma=0.
    \label{eq:supp_constraint}
\end{equation}
Therefore, for each thermodynamic eigenstate, there always exists a gauge such that $\mu_\alpha=\kappa_\alpha$ removes the dominant exponential envelope in every domain and the transformed eigenstate does not exhibit non-Hermitian skin effect(NHSE).

This argument does not apply to discrete boundary states. For such states, even within a single domain, different exponential envelopes may dominate near different boundaries. Therefore, a discrete boundary state does not satisfy Eq.~(\ref{eq:supp_constraint}). Such isolated boundary states are analogous to conventional boundary modes in a single-domain open boundary system~\cite{suppYokomizoNon}. They are not characterized by the generalized Brillouin zone(GBZ) and will not be addressed in the non-Bloch theory developed in this work.

\section{SII.~Topological Origin of Non-Hermitian Skin Effect in Domain-Wall Systems}
\label{sec:supp_topological_origin}

In this section, we give a detailed proof that the non-Hermitian skin effect in domain-wall systems originates from point-gap winding mismatch. We consider a non-Hermitian domain-wall ring (DW ring) composed of $n$ piecewise-translational domains. For a complex energy $E$, the point gap of domain $\alpha$ is open if $\det[h_\alpha(e^{ik})-E]\neq0$ for all $k\in[0,2\pi)$. In this case, the point-gap winding of domain $\alpha$ is well defined as
\begin{equation}
    w_\alpha(E)
    =
    \frac{1}{2\pi i}
    \int_0^{2\pi} dk\,
    \partial_k
    \log\det[h_\alpha(e^{ik})-E].
    \label{eq:supp_winding}
\end{equation}
At the interface $\alpha|\alpha+1$, we define the relative winding number as
\begin{equation}
    \Delta_\alpha(E)
    =
    w_{\alpha+1}(E)-w_\alpha(E).
\end{equation}
We conclude that the point-gap-protected NHSE originates from the positive relative winding at this interface. More precisely, if an eigenstate of DW ring is localized at $\alpha|\alpha+1$, we have $\Delta_\alpha(E)>0$.

We proceed in three steps. First, we present an intuitive argument for a single interface $\alpha|\alpha+1$ formed by two semi-infinite domains whose point gaps at $E$ are both open. Second, for the same semi-infinite geometry, we provide a topological proof of the local interface criterion. Third, we return to an $n$-DW ring and discuss the implication of this local criterion for the closed geometry. We show that point-gap-protected right skin localization at a given interface $\alpha|\alpha+1$ requires a positive winding mismatch $\Delta_\alpha(E)>0$ across that interface. Equivalently, point-gap-protected NHSE is absent when there is no winding mismatch across any interface, namely when all point-gapped domains share the same winding number.

\subsection{An intuitive proof from eigenmode counting}

We first give a physically intuitive proof of the local criterion in a special case.  This argument assumes nondegenerate roots.  A more general Toeplitz-index proof is given afterwards. 
We consider the interface $\alpha|\alpha+1$ as an infinite domain-wall system. Choosing the interface between $x=0$ and $x=1$, domain $\alpha$ occupies the left semi-infinite region $x\leq0$, while domain $\alpha+1$ occupies the right semi-infinite region $x\geq1$. Since this geometry has no external boundaries, the remaining boundary conditions are the matching conditions across the domain wall.

We first rewrite the winding number as an integral in the complex $\beta$ plane.  Let $\beta=e^{ik}$ and denote $f_\alpha(\beta,E)=\det[h_\alpha(\beta)-E]$. In general, $f_\alpha(\beta,E)$ is a Laurent polynomial in $\beta$,
\begin{equation}
    f_\alpha(\beta,E)
    =
   \sum_{n=-s_\alpha}^{p_\alpha}a_{\alpha,n}(E)\beta^n,
\end{equation}
where $s_\alpha$ and $d_\alpha=s_\alpha+p_\alpha$ denote the numbers of poles and roots, respectively. Let $\beta_{\alpha,1}(E),\ldots,\beta_{\alpha,d_\alpha}(E)$ be the roots of $f_\alpha(\beta,E)=0$.  Since $E$ lies in the point gap of the $\alpha$th bulk, no root lies on the unit circle $|\beta|=1$.  Therefore we rewrite the point-gap winding number as
\begin{equation}
    w_\alpha(E)
    =
    \frac{1}{2\pi i}
    \oint_{|\beta|=1}
    d\beta\,
    \partial_\beta\log f_\alpha(\beta,E).
\end{equation}
By the argument principle, we have
\begin{equation}
    w_\alpha(E)
    =
    n_\alpha^<(E)-s_\alpha,
\end{equation}
where $n_\alpha^<(E)$ is the number of roots of $f_\alpha$ inside the unit circle, namely the number of solutions satisfying $|\beta_{\alpha,m}(E)|<1 $. We take the positive spatial direction to be the right direction. Thus $|\beta_{\alpha,m}|<1$ represents a right-decaying mode, whereas $|\beta_{\alpha,m}|>1$ represents a left-decaying mode.  The winding mismatch at the interface $\alpha|\alpha+1$ is therefore
\begin{equation}
    \Delta_\alpha(E)
    =
    w_{\alpha+1}(E)-w_\alpha(E)
    =
    [n_{\alpha+1}^<(E)-s_{\alpha+1}]
    -
    [n_\alpha^<(E)-s_\alpha].
\end{equation}
Hence the relative winding counts the difference between the numbers of right-decaying modes in the two domains, up to the energy-independent shift $s_{\alpha+1}-s_\alpha$.

Choose the local coordinate such that the left side of the interface, $x\le0$, belongs to domain $\alpha$, while the right side, $x\ge1$, belongs to domain $\alpha+1$.  Assuming all roots are nondegenerate, the bulk solutions far from the interface can be written as
\begin{align}
    \psi_x^{(\alpha)}
    &=
    \sum_{m=1}^{d_\alpha}
    a_m\,\beta_{\alpha,m}^{\,x}u_{\alpha,m},
    \qquad x\le0,\\
    \psi_x^{(\alpha+1)}
    &=
    \sum_{m=1}^{d_{\alpha+1}}
    b_m\,\beta_{\alpha+1,m}^{\,x}u_{\alpha+1,m},
    \qquad x\ge1,
\end{align}
where $u_{\alpha,m}$ is the internal eigenvector satisfying $[h_\alpha(\beta_{\alpha,m})-E]u_{\alpha,m}=0$ and $\psi_x^{(\alpha)}$ is in general a vector.

A right eigenstate localized at the interface must decay in both directions, namely as $x\to-\infty$ and as $x\to+\infty$.  Therefore, on the left side only the modes with $|\beta_{\alpha,m}|>1$ are allowed, while on the right side only the modes with $|\beta_{\alpha+1,m}|<1$ are allowed:
\begin{align}
    \psi_x^{(\alpha)}
    &=
    \sum_{|\beta_{\alpha,m}|>1}
    a_m\,\beta_{\alpha,m}^{\,x}u_{\alpha,m},
    \qquad x\le0,\\
    \psi_x^{(\alpha+1)}
    &=
    \sum_{|\beta_{\alpha+1,m}|<1}
    b_m\,\beta_{\alpha+1,m}^{\,x}u_{\alpha+1,m},
    \qquad x\ge1.
\end{align}
Let $n_\alpha^>(E)$ be the number of roots satisfying $|\beta_{\alpha,m}(E)|>1$.  Since there are no roots on the unit circle,
\begin{equation}
    n_\alpha^<(E)+n_\alpha^>(E)=d_\alpha.
\end{equation}
Thus the dimension of the candidate solution space for right eigenstates localized at the domain wall is
\begin{equation}
    D_{\rm sol}^{R}
    =
    n_\alpha^>(E)+n_{\alpha+1}^<(E).
\end{equation}

Next we count the number of constraints imposed by the domain-wall boundary conditions.  For a single-band model, $s_\alpha$ and $d_\alpha-s_\alpha$ are the corresponding left and right hopping range, respectively.  For a model with internal degrees of freedom, one may equivalently regard the internal states as separate sites, so that $s_\alpha$ and $d_\alpha-s_\alpha$ again represent the corresponding hopping distances.  Therefore, the number of independent constraints at the interface $\alpha|\alpha+1$ is
\begin{equation}
    D_{\rm constrain}
    =
    (d_\alpha-s_\alpha)+s_{\alpha+1}.
\end{equation}
The remaining dimension of the right-localized solution space is at least
\begin{align}
    D_{\rm sol}^{R}-D_{\rm constrain}
    &=
    n_\alpha^>(E)+n_{\alpha+1}^<(E)
    -(d_\alpha+s_{\alpha+1}-s_\alpha)
    \notag\\
    &=
    d_\alpha-n_\alpha^<(E)+n_{\alpha+1}^<(E)
    -(d_\alpha+s_{\alpha+1}-s_\alpha)
    \notag\\
    &=
    [n_{\alpha+1}^<(E)-s_{\alpha+1}]
    -
    [n_\alpha^<(E)-s_\alpha]
    \notag\\
    &=
    \Delta_\alpha(E)
    =
    w_{\alpha+1}(E)-w_{\alpha}(E).
\end{align}
Therefore, when $\Delta_\alpha(E)>0$, the candidate degrees of freedom for localized eigenstates outnumber the constraints, and NHSE must exist.

\subsection{General proof from the Toeplitz index theorem}

The above proof is physically intuitive but not fully general.  We now give a more general proof based on the Toeplitz index theorem~\cite{suppTrefethenSpectra,suppBottcherSpectral}.  The strategy is the same as that used in Ref.~\cite{suppOkumaTopological} for open-boundary systems. 

Following the semi-infinite boundary condition (SIBC) construction, we first cut the hopping terms at the interface $\alpha|\alpha+1$.  The system is then separated into two decoupled semi-infinite chains.  Domain $\alpha$ lies on
the left of the interface, with coordinate $x\le0$; it is a left semi-infinite chain terminated at $x=0$.  Domain $\alpha+1$ lies on the right of the interface, with coordinate $x\ge1$; it is a right semi-infinite chain terminated at $x=1$.  We denote their real-space Hamiltonians by $H_{\alpha}^{\rm SIBC,-}$ and $H_{\alpha+1}^{\rm SIBC,+}$, respectively, where the superscripts $-$ and $+$ label the left and right semi-infinite geometries.  The reference Hamiltonian after cutting the interface is
\begin{equation}
    H_{\alpha|\alpha+1}^{(0)}
    =
    H_{\alpha}^{\rm SIBC,-}
    \oplus
    H_{\alpha+1}^{\rm SIBC,+}.
\end{equation}

Inside each domain, the hopping matrix depends only on the relative distance between two sites.  Hence, in real space, all matrix elements along the same diagonal are identical; for models with internal degrees of freedom, the same statement holds for matrix blocks.  On a semi-infinite lattice, a Toeplitz operator is represented by a matrix with entries constant along each diagonal~\cite{suppTrefethenSpectra,suppBottcherSpectral}. Therefore $H_{\alpha}^{\rm SIBC,-}$ and $H_{\alpha+1}^{\rm SIBC,+}$ are Toeplitz operators.

Assume that $E$ lies in the point gap of the $\alpha$th bulk:
\begin{equation}
    \det[h_\alpha(e^{ik})-E]\neq0,   \quad \forall k\in[0,2\pi).
\end{equation}
This means that the matrix function $h_\alpha(e^{ik})-E$ is invertible on the unit circle $|\beta|=1$.  Therefore the semi-infinite Toeplitz operator $H_{\alpha}^{\rm SIBC,-}-E$ is Fredholm~\cite{suppBottcherSilbermann,suppBottcherSpectral}.  Similarly, since $E$ also lies in the point gap of the $(\alpha+1)$th bulk, $H_{\alpha+1}^{\rm SIBC,+}-E$ is Fredholm.  For a Fredholm operator $F$, the kernel $\ker F$ and cokernel $\mathrm{coker}\,F$ are finite-dimensional.  Moreover, $\mathrm{coker}\,F$ is isomorphic to $\ker F^\dagger$.  Thus its Fredholm index is a well-defined integer
\begin{equation}
    \operatorname{ind}F
    =
    \dim\ker F-\dim\mathrm{coker}F
    =
    \dim\ker F-\dim\ker F^\dagger .
\end{equation}
We define
\begin{equation}
    F_\alpha^-(E)=H_{\alpha}^{\rm SIBC,-}-E,
    \qquad
    F_{\alpha+1}^+(E)=H_{\alpha+1}^{\rm SIBC,+}-E,
\end{equation}
and
\begin{equation}
    F_{\alpha|\alpha+1}^{(0)}(E)
    =
    F_\alpha^-(E)\oplus F_{\alpha+1}^+(E).
\end{equation}

The Toeplitz index theorem gives~\cite{suppTrefethenSpectra,suppBottcherSpectral}
\begin{equation}
    \operatorname{ind}F_\alpha^-(E)=-w_\alpha(E),
    \qquad
    \operatorname{ind}F_{\alpha+1}^+(E)=w_{\alpha+1}(E).
\end{equation}
The opposite signs arise from the opposite orientations of the left and right
semi-infinite spaces.  Hence the Toeplitz index of the reference operator is
\begin{align}
    \operatorname{ind}F_{\alpha|\alpha+1}^{(0)}(E)
    &=
    \operatorname{ind}F_\alpha^-(E)
    +
    \operatorname{ind}F_{\alpha+1}^+(E)
    \notag\\
    &=
    w_{\alpha+1}(E)-w_\alpha(E)
    =
    \Delta_\alpha(E).
\end{align}

We now return to the actual infinite domain-wall Hamiltonian $H_{\alpha|\alpha+1}$.  Its difference from the reference Hamiltonian $H_{\alpha|\alpha+1}^{(0)}$ is localized within finitely many sites near the domain wall.  This difference is therefore a finite-rank perturbation, hence a compact perturbation.  Since the Fredholm index is invariant under compact perturbations~\cite{suppKatoPerturbation},
\begin{equation}
    \operatorname{ind}F_{\alpha|\alpha+1}(E)
    =
    \operatorname{ind}F_{\alpha|\alpha+1}^{(0)}(E)
    =
    \Delta_\alpha(E),
\end{equation}
where $F_{\alpha|\alpha+1}(E)=H_{\alpha|\alpha+1}-E$.

Finally, we explain why a nonzero index implies a skin state at the domain wall.  By the definition of the Fredholm index,
\begin{equation}
    \operatorname{ind}F_{\alpha|\alpha+1}(E)
    =
    \Delta_\alpha(E)
    =
    \dim\ker F_{\alpha|\alpha+1}(E)
    -
    \dim\ker F_{\alpha|\alpha+1}^{\dagger}(E).
\end{equation}
If $\Delta_\alpha(E)>0$, then
\begin{equation}
    \dim\ker F_{\alpha|\alpha+1}(E)
    =
    \Delta_\alpha(E)
    +
    \dim\ker F_{\alpha|\alpha+1}^{\dagger}(E)
    \ge
    \Delta_\alpha(E)>0.
\end{equation}
Hence there exists a nonzero right eigenstate $\psi$ satisfying $(H_{\alpha|\alpha+1}-E)\psi=0$. If $\Delta_\alpha(E)<0$, the same identity implies
\begin{equation}
    \dim\ker F_{\alpha|\alpha+1}^{\dagger}(E)>0,
\end{equation}
and therefore there exists a nonzero left eigenstate $\phi$ satisfying
\begin{equation}
    (H_{\alpha|\alpha+1}^{\dagger}-E^*)\phi=0.
\end{equation}

Far away from the interface, the system reduces to the two translationally invariant bulks.  Therefore the states in the kernel can only be composed of the corresponding non-Bloch bulk solutions.  Since $E$ lies in the point gaps of both bulks, the characteristic equations of the two bulks have no roots on the unit circle.  Equivalently, there are no ordinary propagating plane waves with $|\beta|=1$.  On the other hand, $\ker F_{\alpha|\alpha+1}(E)$ is taken in the Hilbert space of the two-sided infinite chain, so its states must be square-summable both as $x\to-\infty$ and as $x\to+\infty$.  Thus, in domain $\alpha$, only modes decaying toward $x\to-\infty$ are allowed, which means they are localized near the interface; in domain $\alpha+1$, only modes decaying toward $x\to+\infty$ are allowed, again localized near the interface. Therefore the right or left eigenstates protected by the nonzero index are necessarily interface skin states near the domain wall.  This proves that a nonzero winding mismatch is the topological origin of the NHSE in a domain-wall system. Finally, for a local interface $\alpha|\alpha+1$, we conclude that

\begin{align}
    &\Delta_\alpha(E)>0
   \,\Leftrightarrow\,
    \text{a point-gap-protected right eigenstate localized at } \alpha|\alpha+1, \\
    &\Delta_\alpha(E)<0
   \,\Leftrightarrow\,
    \text{a point-gap-protected left eigenstate localized at } \alpha|\alpha+1, \\
    &\Delta_\alpha(E)= 0
   \,\Leftrightarrow\,
    \text{no topologically protected eigenstate localized at } \alpha|\alpha+1 .
\end{align}

\subsection{From a local interface to an $n$-domain-wall ring}

We now return to a closed ring with $n$ domains. If the point gaps of some domains close at a given energy $E$, their characteristic equations admit unit-modulus solutions, $|\beta_{\alpha,m}|=1$, corresponding to Bloch-wave modes. In the thermodynamic limit, these modes dominate the wave functions in the bulk of the corresponding domains, rather than producing an exponentially accumulating skin profile. We therefore focus below on the case in which all domains are point-gapped at $E$, so that all winding numbers and relative windings are well defined.

Our conclusion is the following. If a right eigenstate of the DW ring exhibits skin localization near $\alpha|\alpha+1$ in the thermodynamic limit, the relative winding across this interface must be positive, namely,
\begin{equation}
    \Delta_\alpha(E)
    =
    w_{\alpha+1}(E)-w_\alpha(E)>0.
    \label{eq:supp_ring_necessary}
\end{equation}
Likewise, a left eigenstate localized near the same interface requires $\Delta_\alpha(E)<0$.

To see this, we recall that, for a single interface with both domains being semi-infinite, the local criterion established above is both necessary and sufficient: a positive (negative) winding mismatch gives a point-gap-protected right (left) skin mode at the interface, whereas a vanishing mismatch protects no skin mode. Now suppose that an exact DW-ring eigenstate is exponentially localized near $\alpha|\alpha+1$. As all domain lengths tend to infinity, the other domain walls become macroscopically far away. We may therefore truncate the eigenstate at distances much larger than its localization length but much smaller than the separation between domain walls. The removed tails are exponentially small, and the truncation error is confined to the cuts and vanishes in the thermodynamic limit. The remaining wave function consequently converges to a normalizable eigenstate of the corresponding two-sided infinite-interface problem, in which the two adjoining domains extend to semi-infinity. Applying the local criterion then gives Eq.~(\ref{eq:supp_ring_necessary}).

For a single infinite interface, a nonzero winding mismatch is sufficient to produce the corresponding local skin mode. In a closed DW ring, however, a nonzero $\Delta_\alpha(E)$ does not by itself require a given global eigenstate to localize at $\alpha|\alpha+1$, because its wave function must simultaneously satisfy the non-Hermitian boundary conditions. For example, consider a three-domain ring and an eigenenergy $E$ satisfying $w_3(E)>w_2(E)>w_1(E)$. In this case, we have
\begin{equation}
    \Delta_1(E)=w_2(E)-w_1(E)>0,
    \qquad
    \Delta_2(E)=w_3(E)-w_2(E)>0.
\end{equation}
The two interfaces $1|2$ and $2|3$ can therefore support right skin modes when considered separately. For the thermodynamic eigenstate considered here, however, the wave function in each domain has a single dominant exponential envelope in the thermodynamic limit. In domain $2$, its envelope is therefore proportional to $|\beta_2|^x$. It cannot decay away from both $1|2$ and $2|3$ simultaneously, as this would require two independent dominant modes localized at the two ends of domain $2$. The global boundary conditions consequently select at most one of these two interfaces as the localization center of a given eigenstate.
Thus, although both positive winding mismatches identify possible topological origins of skin localization, the relative windings alone cannot determine which interface is ultimately selected in the closed ring.

We can nevertheless obtain a general necessary condition. If a DW-ring eigenstate exhibits the NHSE, it must accumulate near at least one domain wall, and the winding mismatch across this interface must be nonzero. Therefore, for a given eigenenergy $E$, we have
\begin{equation}
    \mathrm{NHSE}
    \quad\Longrightarrow\quad
    \Delta_\alpha(E)\neq0
    \quad\text{for at least one }\alpha .
    \label{eq:supp_ring_origin}
\end{equation}
Since all relative windings vanish if and only if all domains share the same winding number,
\begin{equation}
    w_1(E)=w_2(E)=\cdots=w_n(E),
    \label{eq:supp_commonwregion}
\end{equation}
the common-winding condition is sufficient to exclude the point-gap-protected NHSE. Hence, relative point-gap topology is the only origin of the NHSE in a DW-ring system.

\section{SIII.~Domain-Wall-Ring GBZ From Boundary Conditions}
\label{sec:supp_matching_matrix}

In this section, we present a rigorous derivation of GBZ equations from the DW ring boundary conditions. We consider the most general DW ring systems where each domain is multiband and the associated characteristic Laurent series has different orders,
\begin{equation}
    f_\alpha(\beta,E)=\det [h_\alpha(\beta)-E]=\sum_{n=-s_\alpha}^{p_\alpha} a_{\alpha,n}(E)\beta^n.
    \label{ep:supp_Laurent}
\end{equation}
For domain $\alpha$, the number of roots for the characteristic equation is $d_\alpha=s_\alpha+p_\alpha$.

\subsection{Derivation of boundary equations}

We now derive the boundary equations of the DW ring from the real space wave functions and the DW ring boundary conditions.  In the $\alpha$th domain, the internal eigenvector $u_{\alpha,m}$ associated with
$\beta_{\alpha,m}$ satisfies
\begin{equation}
    [h_\alpha(\beta_{\alpha,m})-E]u_{\alpha,m}=0.
\end{equation}
We choose $x=0,1,\ldots,N_\alpha-1$ to be a local coordinate inside the $\alpha$th domain. We assume, for simplicity, that no degenerate roots are present. Then the bulk solution in this domain can be written as
\begin{equation}
    \psi_x^{(\alpha)}
    =
    \sum_{m=1}^{d_\alpha}
    c_{\alpha,m}
    \beta_{\alpha,m}^{x}
    u_{\alpha,m}.
\end{equation}
We collect the coefficients into the column vector $C_\alpha$
\begin{equation}
    C_\alpha
    =
    (c_{\alpha,1},c_{\alpha,2},\ldots,c_{\alpha,d_\alpha})^T .
\end{equation}

We first write the wave function near the right end of the $\alpha$th domain.  Because the hopping range is finite, the boundary condition at the interface $\alpha|\alpha+1$ only involves finite sites near the right end of domain $\alpha$. The right side of domain $\alpha$ contributes $p_\alpha$ independent effective variables to the boundary equation~\cite{suppYokomizoNon}.  For a single-band model, these variables can be identified with the $p_\alpha$ sites near the right boundary.  For a multiband model, they may involve internal components and their linear combinations.

We label these effective variables by $j=1,\ldots,p_\alpha$. The $j$th variable entering the boundary condition can be written as
\begin{equation}
    \phi_{\alpha,j}^{R}
    =
    \sum_{m=1}^{d_\alpha}
    c_{\alpha,m}
    \beta_{\alpha,m}^{N_\alpha}
    v_{\alpha, j,m}^{R}(E).
\end{equation}
Here $v_{\alpha, j,m}^{R}(E)$ is a finite boundary coefficient.  It contains the internal-state components and finite factors of $\beta_{\alpha,m}$ and no factor exponentially dependent on $N_\alpha$. For single-band models, $v_{\alpha, j ,m}^R(E)=\beta_{\alpha,m}^{-j}(E)$, such that $\phi_{\alpha,j}^R$ is the wave function at $N_\alpha-j$. For multi-band, $v_{\alpha,j,m}^R(E)$ contains both finite factors of $\beta_{\alpha,m}$ and projection coefficients of eigenvector $u_{\alpha,m}$.

We collect all wave-function components of domain $\alpha$ that enter the boundary condition into the vector
\begin{equation}
    \Psi_\alpha^R
    =
    \begin{pmatrix}
        \phi_{\alpha,1}^{R}\\
        \phi_{\alpha,2}^{R}\\
        \vdots\\
        \phi_{\alpha,p_\alpha}^{R}
    \end{pmatrix}.
\end{equation}

For each $\phi_{\alpha,j}^{R}$, we can further decompose it into three parts:
\begin{align}
    \phi_{\alpha,j}^{R}
   & =
    \sum_{m=1}^{d_\alpha}
    c_{\alpha,m}
    \beta_{\alpha,m}^{N_\alpha}
    v_{\alpha, j,m}^{R}.
    \notag\\
    &=
    \begin{pmatrix}
        v_{\alpha, j,1}^R &
        v_{\alpha ,j,2}^R &
        \cdots &
        v_{\alpha, j,d_\alpha}^R
    \end{pmatrix}
    \begin{pmatrix}
        \beta_{\alpha,1}^{N_\alpha} & & & \\
        & \beta_{\alpha,2}^{N_\alpha} & & \\
        & & \ddots & \\
        & & & \beta_{\alpha,d_\alpha}^{N_\alpha}
    \end{pmatrix}
    \begin{pmatrix}
        c_{\alpha,1}\\
        c_{\alpha,2}\\
        \vdots\\
        c_{\alpha,d_\alpha}
    \end{pmatrix}
    \notag\\
    &=
    U_{\alpha,j}^{R}D_\alpha C_\alpha .
\end{align}
Here
\begin{equation}
    U_{\alpha,j}^{R}
    =
       \begin{pmatrix}
        v_{\alpha, j,1}^R &
        v_{\alpha ,j,2}^R &
        \cdots &
        v_{\alpha ,j,d_\alpha}^R
    \end{pmatrix},
\end{equation}
and $D_\alpha$ is a $d_\alpha\times d_\alpha$ diagonal matrix,
\begin{equation}
    D_\alpha
    =
    \mathrm{diag}
    \left(
        \beta_{\alpha,1}^{N_\alpha},
        \beta_{\alpha,2}^{N_\alpha},
        \ldots,
        \beta_{\alpha,d_\alpha}^{N_\alpha}
    \right).
\end{equation}
Therefore,
\begin{align}
    \Psi_\alpha^R
    &=
    \begin{pmatrix}
        U_{\alpha,1}^{R}D_\alpha C_\alpha\\
        U_{\alpha,2}^{R}D_\alpha C_\alpha\\
        \vdots\\
        U_{\alpha,p_\alpha}^{R}D_\alpha C_\alpha
    \end{pmatrix}
    \notag\\
    &=
    \begin{pmatrix}
        U_{\alpha,1}^{R}\\
        U_{\alpha,2}^{R}\\
        \vdots\\
        U_{\alpha,p_\alpha}^{R}
    \end{pmatrix}
    D_\alpha C_\alpha
    =
    U_\alpha^R D_\alpha C_\alpha ,
\end{align}
where $U_\alpha^R(E)$ is a $p_\alpha\times d_\alpha$ matrix.

Next, for domain $\alpha+1$, the formal solution is
\begin{equation}
    \psi_{x}^{(\alpha+1)}
    =
    \sum_{m=1}^{d_{\alpha+1}}
    c_{\alpha+1,m}\beta_{\alpha+1,m}^x
    u_{\alpha+1,m},
\end{equation}
where $x=0,1,\ldots,N_{\alpha+1}-1$ is the local coordinate. Similarly, we consider the independent variables near the left end of the $(\alpha+1)$th domain that enter the same interface boundary condition. Their number is $s_{\alpha+1}$, and we denote the corresponding wave function vector by $\Psi_{\alpha+1}^L$. The left-boundary wave-function vector are collected as
\begin{equation}
    \Psi_{\alpha+1}^L
    =
    \begin{pmatrix}
        \phi_{\alpha+1,1}^{L}\\
       \phi_{\alpha+1,2}^{L}\\
        \vdots\\
       \phi_{\alpha+1,s_{\alpha+1}}^{L}
    \end{pmatrix}.
\end{equation}
For each component,
\begin{align}
    \phi_{\alpha+1,j}^L&=\sum_{m=1}^{d_{\alpha+1}}
    c_{\alpha+1,m}    v_{\alpha+1,j,m}^L
    \notag\\
    &=
    \begin{pmatrix}
        v_{\alpha+1,j,1}^L &
        v_{\alpha+1,j,2}^L&
        \cdots &
        v_{\alpha+1,j,d_{\alpha+1}}^L
    \end{pmatrix}
    \begin{pmatrix}
        c_{\alpha+1,1}\\
        c_{\alpha+1,2}\\
        \vdots\\
        c_{\alpha+1,d_{\alpha+1}}
    \end{pmatrix}
    \notag\\
    &=
    U_{\alpha+1,j}^{L}C_{\alpha+1}.
\end{align}
As in domain $\alpha$, for single-band models, $v_{\alpha+1,j,m}=\beta_{\alpha+1,m}^{j-1}$. Thus
\begin{align}
    \Psi_{\alpha+1}^{L}
    &=
    \begin{pmatrix}
        U_{\alpha+1,1}^{L}C_{\alpha+1}\\
        U_{\alpha+1,2}^{L}C_{\alpha+1}\\
        \vdots\\
        U_{\alpha+1,s_{\alpha+1}}^{L}C_{\alpha+1}
    \end{pmatrix}
    \notag\\
    &=
    \begin{pmatrix}
        U_{\alpha+1,1}^{L}\\
        U_{\alpha+1,2}^{L}\\
        \vdots\\
        U_{\alpha+1,s_{\alpha+1}}^{L}
    \end{pmatrix}
    C_{\alpha+1}
    =
    U_{\alpha+1}^{L}C_{\alpha+1},
\end{align}
where $U_{\alpha+1}^L$ is a $s_{\alpha+1}\times d_{\alpha+1}$ matrix.

The real-space boundary condition at the interface $\alpha|\alpha+1$ can now be written as
\begin{equation}
    \widetilde L_\alpha(E)\Psi_\alpha^R
    +
    \widetilde G_{\alpha+1}(E)\Psi_{\alpha+1}^L
    =
    0 .
\end{equation}
Here $\widetilde L_\alpha(E)$ and $\widetilde G_{\alpha+1}(E)$ depend only on the energy and on the local hopping coefficients near the interface. Since there are in total $(p_\alpha+s_{\alpha+1})$ independent boundary equations, the dimensions of $\widetilde L_\alpha(E)$ and $\widetilde G_{\alpha+1}(E)$ are $(p_\alpha+s_{\alpha+1})\times p_\alpha$ and $(p_\alpha+s_{\alpha+1})\times s_{\alpha+1}$, respectively. Substituting the expressions for $\Psi_\alpha^R$ and $\Psi_{\alpha+1}^L$, we obtain
\begin{equation}
    \widetilde L_\alpha(E)U_\alpha^R(E)D_\alpha(E)C_\alpha
    +
    \widetilde G_{\alpha+1}(E)U_{\alpha+1}^L(E)C_{\alpha+1}
    =
    0.
\end{equation}
This is the boundary equation at the interface $\alpha|\alpha+1$.  We use the convention $C_{n+1}\equiv C_1$.

Collecting all domain-wall boundary equations, we obtain
\begin{equation}
    M(E)
    \begin{pmatrix}
        C_1\\
        C_2\\
        \vdots\\
        C_n
    \end{pmatrix}
    =
    0.
\end{equation}
The coefficient matrix $M(E)$ has the cyclic block form
\begin{equation}
    M(E)
    =
    \begin{pmatrix}
        \widetilde L_1U_1^RD_1
        &
        \widetilde G_2U_2^L
        &
        0
        &
        \cdots
        &
        0
        &
        0\\
        0
        &
        \widetilde L_2U_2^RD_2
        &
        \widetilde G_3U_3^L
        &
        \cdots
        &
        0
        &
        0\\
        \vdots
        &
        \vdots
        &
        \vdots
        &
        \ddots
        &
        \vdots
        &
        \vdots\\
        0
        &
        0
        &
        0
        &
        \cdots
        &
        \widetilde L_{n-1}U_{n-1}^RD_{n-1}
        &
        \widetilde G_nU_n^L\\
        \widetilde G_1U_1^L
        &
        0
        &
        0
        &
        \cdots
        &
        0
        &
        \widetilde L_nU_n^RD_n
    \end{pmatrix}.
\end{equation}
For compactness, we define $L_\alpha=\widetilde L_\alpha U_\alpha^R,\, G_{\alpha+1}=\widetilde G_{\alpha+1} U_{\alpha+1}^L$, their dimensions are $(p_\alpha+s_{\alpha+1})\times d_\alpha$ and $(p_\alpha+s_{\alpha+1})\times d_{\alpha+1}$, respectively. Then
\begin{equation}
    M(E)
    =
    \begin{pmatrix}
        L_1D_1 & G_2 & 0 & \cdots & 0 & 0\\
        0 & L_2D_2 &G_3 & \cdots & 0 & 0\\
        \vdots & \vdots & \vdots & \ddots & \vdots & \vdots\\
        0 & 0 & 0 & \cdots & L_{n-1}D_{n-1} & G_n\\
        G_1 & 0 & 0 & \cdots & 0 & L_nD_n
    \end{pmatrix}.
    \label{eq:supp_M}
\end{equation}
All $\beta_{\alpha,m}$ dependence in $U_\alpha^{L,R}$, and hence in $L_\alpha$ and $G_\alpha$, are finite powers.  Therefore $L_\alpha$ and $G_\alpha$ depend on the local interface structure, the internal eigenvectors, and finite propagation factors, but they do not scale exponentially with $N_\alpha$.  By contrast, the diagonal matrix $D_\alpha$ contains the factors $\beta_{\alpha,m}^{N_\alpha}$.  In the thermodynamic limit, the exponential length dependence is therefore entirely controlled by the matrices $D_\alpha$. Since each $D_\alpha$ is diagonal, every entry of $L_\alpha D_\alpha$ carries a length-dependent factor $\beta_{\alpha,m}^{N_\alpha}$.

The total numbers of rows and columns of matrix $M(E)$ are $\sum_{\alpha} (p_\alpha+s_{\alpha+1})=\sum_\alpha d_\alpha $ and $\sum_\alpha d_\alpha$, respectively. Therefore, $M(E)$ is a square matrix, and the condition for a nonzero solution to exist is that its determinant $\det M(E)$ vanishes.

We now expand $\det M(E)$. In a determinant expansion, every nonzero term can be viewed as choosing one nonzero matrix element from each column, with each row also chosen exactly once.  We use this selection process to determine the possible exponential factors.

For the $\alpha$th block column $\begin{pmatrix}
    0& \cdots& 0&G_\alpha & L_\alpha D_\alpha & 0 \cdots 0
\end{pmatrix}^T$ in Eq.~(\ref{eq:supp_M}), we denote the set of columns selected from the $(p_\alpha+s_{\alpha+1})\times d_\alpha$ block $L_\alpha D_\alpha$ as $I_\alpha$, which satisfies
\begin{equation}
    I_\alpha\subset\{1,2,\ldots,d_\alpha\},
    \qquad
    |I_\alpha|=k_\alpha.
\end{equation}
Then the full length-dependent factor contributed by this domain is $\prod_{m\in I_\alpha}\beta_{\alpha,m}^{N_\alpha}$. The remaining $d_\alpha-k_\alpha$ columns of the same block column must be selected from the cyclic block $G_\alpha$.

A nonzero determinant term must cover all rows and columns exactly once.  The block row associated with the interface $\alpha|\alpha+1$ receives columns from $L_\alpha D_\alpha$ and from $G_{\alpha+1}$.  If $k_\alpha$ columns are chosen from $L_\alpha D_\alpha$, then the number of columns that must be chosen from $G_{\alpha+1}$ is $(p_\alpha+s_{\alpha+1})-k_\alpha$.  But the columns of $G_{\alpha+1}$ belong to the $(\alpha+1)$th block column.  Therefore the number of columns left for $L_{\alpha+1}D_{\alpha+1}$ is 
\begin{equation}
    k_{\alpha+1}=d_{\alpha+1}-(p_\alpha+s_{\alpha+1})+k_\alpha=p_{\alpha+1}-p_\alpha+k_\alpha.
\end{equation}
Hence we have $p_{\alpha+1}-k_{\alpha+1}=p_\alpha-k_\alpha$, repeating this argument around the closed ring, we conclude for every domain, we have
\begin{equation}
    k_\alpha=p_\alpha-w,
\end{equation}
where $w$ is an integer. Since $0\le k_\alpha\le d_\alpha$, $w$ ranges from $-\min_\alpha s_\alpha$ to $\min_\alpha p_\alpha$. 
Note that, as $\mu_\alpha$ varies, the point-gap winding number in domain $\alpha$ takes the values $w_\alpha(\mu_\alpha;E)=-s_\alpha,-s_\alpha+1,\ldots,p_\alpha$.
Consequently, the common winding number shared by all domains can take integer values only from $w_-=-\min_\alpha s_\alpha$ to $w_+=\min_\alpha p_\alpha$. 
We can therefore identify the allowed range of $w$ as the intersection of the winding-number ranges of all domains.

All factors that do not scale exponentially with the domain lengths can be collected into a coefficient $A_{w;I_1,\ldots,I_n}(E)$.  The determinant then takes the form
\begin{equation}
    \det M(E)
    =
    \sum_{w=-\min_\alpha s_\alpha}^{\min_\alpha p_\alpha}
    \sum_{\substack{
        I_1,\ldots,I_n\\
        |I_\alpha|=p_\alpha-w
    }}
    A_{w;I_1,\ldots,I_n}(E)
    \prod_{\alpha=1}^{n}
    \left(
        \prod_{m\in I_\alpha}
        \beta_{\alpha,m}(E)
    \right)^{N_\alpha}.
    \label{eq:supp_detM_expansion}
\end{equation}
Here $I_\alpha$ is the set of root labels selected in the $\alpha$th domain, $I_\alpha\subset\{1,2,\ldots,d_\alpha\}$, and $|I_\alpha|=p_\alpha-w$ means that $p_\alpha-w$ roots are selected in each domain.  Thus, in a given term, every domain contributes the number $p_\alpha-w$ of roots, weighted by its own length, and $w$ ranges from $-\min_\alpha s_\alpha$ to $\min_\alpha p_\alpha$.

Therefore the length dependence of every nonzero determinant term must be of the form $\prod_{\alpha=1}^{n}
    \left(
        \prod_{m\in I_\alpha}
        \beta_{\alpha,m}
    \right)^{N_\alpha}$.

Define the set of all possible exponential factors
\begin{equation}
    \eta
    =
    \left\{
    \prod_{\alpha=1}^{n}
    \left(
        \prod_{m\in I_\alpha}
        \beta_{\alpha,m}(E)
    \right)^{N_\alpha}
    \,\middle|\,
    w=w_-,w_-+1,\cdots w_+,\;
    |I_\alpha|=p_\alpha-w,\;
    I_\alpha\subset\{1,2,\ldots,d_\alpha\}
    \right\},
\end{equation}
where $w_-=-\min_\alpha s_\alpha,\,w_+=\min _\alpha p_\alpha$. In the thermodynamic limit, if one element of $\eta$ has a modulus strictly larger than all the others, and if its coefficient $A_{w;I_1,\ldots,I_n}(E)$ is nonzero, this term exponentially dominates $\det M(E)$.  It cannot be cancelled by the other terms.  Therefore the GBZ condition is that the largest modulus in the set $\eta$ is attained by at least two terms. 

In contrast, as in ordinary open chains~\cite{suppYokomizoNon}, discrete boundary solutions may arise when the coefficient of a single dominant term $A_{w;I_1,\ldots,I_n}(E)$ vanishes.  Such a solution is not produced by the condition above and therefore is not part of the continuum spectrum considered here.

\subsection{Thermodynamic reduction of the GBZ condition}
\label{sec:supp_gbz_simplification}

We now simplify the equal-dominance condition.  Since the GBZ condition only depends on the moduli of $\beta_{\alpha,m}$, we work with $\mu_{\alpha,m}=\log|\beta_{\alpha,m}|$ and define
\begin{equation}
    \zeta
    =
    \left\{
    \sum_{\alpha=1}^{n}
    \sum_{m\in I_\alpha}
    r_\alpha\mu_{\alpha,m}(E)
    \,\middle|\,
    w=w_-,w_-+1,\cdots,w_+,\;
    |I_\alpha|=p_\alpha-w,\;
    I_\alpha\subset\{1,2,\ldots,d_\alpha\}
    \right\},
\end{equation}
where $w_-=-\min_\alpha(s_\alpha),\,w_+=\min_\alpha p_\alpha$ and $r_\alpha=N_\alpha/\sum_\gamma N_\gamma$. The GBZ condition is that the two largest elements of $\zeta$ are equal. For each domain, the roots are ordered by modulus,
\begin{equation}
    \mu_{\alpha,1}
    \le
    \mu_{\alpha,2}
    \le
    \cdots
    \le
    \mu_{\alpha,d_\alpha},
    \qquad
    \forall \alpha .
    \label{eq:supp_mu_ordering}
\end{equation}

As in the determinant expansion, we decompose the possible exponential orders into common-winding sectors.  For a fixed $w\in\{w_-,w_-+1,\ldots,w_+\}$, we define
\begin{equation}
    \zeta_w
    =
    \left\{
    \sum_{\alpha=1}^{n}
    \sum_{m\in I_\alpha}
    r_\alpha\mu_{\alpha,m}(E)
    \,\middle|\,
    |I_\alpha|=p_\alpha-w,\;
    I_\alpha\subset\{1,2,\ldots,d_\alpha\}
    \right\},
    \label{eq:supp_zeta_w}
\end{equation}
The full set of exponential orders is therefore
\begin{equation}
    \zeta
    =
    \bigcup_{w=w_-}^{w_+}\zeta_w .
    \label{eq:supp_zeta_decomposition}
\end{equation}

We first determine the largest element in each sector.  In the $w$ sector, domain $\alpha$ contributes $p_\alpha-w$ roots.  Because of the ordering in Eq.~(\ref{eq:supp_mu_ordering}), its largest contribution is obtained by selecting
\begin{equation}
    \mu_{\alpha,s_\alpha+w+1},
    \mu_{\alpha,s_\alpha+w+2},
    \ldots,
    \mu_{\alpha,d_\alpha}.
\end{equation}
The maximum of $\zeta_w$ is thus
\begin{equation}
    \Lambda_w(E)
    =
    \sum_{\alpha=1}^{n}
    \sum_{m=s_\alpha+w+1}^{d_\alpha}
    r_\alpha\mu_{\alpha,m}(E),
    \qquad
    w=w_-,w_-+1,\ldots,w_+,
    \label{eq:supp_Lambda_w}
\end{equation}
where an empty sum is understood to be zero.  We denote the largest exponential order among all sectors as $\Lambda_*(E)=\max_{w_-\le w\le w_+}\Lambda_w(E)$. To determine which sector reaches $\Lambda_*$, we further introduce the difference between two adjacent sector maxima,
\begin{align}
    \lambda_w(E)
    & =
    \Lambda_{w-1}(E)-\Lambda_w(E)
    \notag\\
    & =
    \sum_{\alpha=1}^{n}
    r_\alpha\mu_{\alpha,s_\alpha+w}(E),
    \qquad
    w=w_-+1,\ldots,w_+ .
    \label{eq:supp_lambda_w}
\end{align}
The root ordering gives
\begin{equation}
    \lambda_{w+1}-\lambda_w
    =
    \sum_{\alpha=1}^{n}
    r_\alpha
    \left(
        \mu_{\alpha,s_\alpha+w+1}
        -
        \mu_{\alpha,s_\alpha+w}
    \right)
    \ge0,
    \label{eq:supp_lambda_monotonic}
\end{equation}
so $\lambda_w$ is non-decreasing with $w$.  For convenience, we set $ \lambda_{w_-}\equiv-\infty,
    \lambda_{w_++1}\equiv+\infty .$ A sector $w$ is maximal precisely when
\begin{equation}
    \lambda_w(E)\le0\le\lambda_{w+1}(E).
    \label{eq:supp_maximal_w_sector}
\end{equation}
Indeed, the first inequality means that the sector maxima do not decrease as one approaches $w$ from below, while the second means that they do not increase after $w$.  Together with the monotonicity of $\lambda_w$, these inequalities imply $\Lambda_w=\Lambda_*$.

The largest element of $\zeta$ can now be degenerate in two different ways.  The first possibility is an internal degeneracy within a maximal $w$ sector. The maximal term in $\zeta_w$ selects the $p_\alpha-w$ largest roots of each domain. Another term in the same sector can have the same exponential order only if, in at least one domain, the largest unselected root and the smallest selected root have the same modulus. Thus, when the selection boundary lies within the ordered roots of a domain, the internal-degeneracy condition is
\begin{equation}
    \mu_{\alpha,s_\alpha+w}(E)
    =
    \mu_{\alpha,s_\alpha+w+1}(E),
    \qquad
    1\le s_\alpha+w\le d_\alpha-1.
    \label{eq:supp_internal_degeneracy}
\end{equation}
This equality contributes to the GBZ only when the corresponding sector is maximal, as required by Eq.~(\ref{eq:supp_maximal_w_sector}).

The second possibility is a degeneracy between two adjacent sectors. By Eq.~(\ref{eq:supp_lambda_w}), the maxima of sectors $w-1$ and $w$ are equal if and only if
\begin{equation}
    \lambda_w(E)=0.
    \label{eq:supp_adjacent_degeneracy}
\end{equation}
No additional maximal-sector condition is needed.  Since $\lambda_w$ is non-decreasing, $\lambda_w=0$ implies that the sector maxima increase up to $w-1$ and decrease after $w$; hence $\Lambda_{w-1}=\Lambda_w=\Lambda_*$.

Combining the two possibilities, the thermodynamic GBZ condition is
\begin{align}
    &\mathrm{Case\ I:}\qquad\,\,
    \exists w\in\{w_-,w_-+1,\cdots,w_+\},
    \qquad
    \lambda_w(E)\le0\le\lambda_{w+1}(E),
    \qquad
    \exists\alpha:
    \mu_{\alpha,s_\alpha+w}(E)
    =
    \mu_{\alpha,s_\alpha+w+1}(E),
    \label{eq:supp_GBZ_case1}
    \\
    &\mathrm{Case\ II:}\qquad
    \exists w\in\{w_-+1,\ldots,w_+\},
    \qquad
    \lambda_w(E)=0 ,
    \label{eq:supp_GBZ_case2}
\end{align}
Case I is an internal degeneracy within a maximal common-winding sector, whereas Case II is a degeneracy between two adjacent common-winding sectors.

The two cases have distinct physical meanings.  In Case I, the two dominant determinant terms differ only by exchanging $\beta_{\alpha,s_\alpha+w}$ and $\beta_{\alpha,s_\alpha+w+1}$ in one domain.  Their equal moduli allow the two corresponding non-Bloch components to coexist at the same exponential order.  Their superposition produces a standing-wave structure in domain $\alpha$.  Case I is therefore the domain-wall generalization of the conventional open-boundary-GBZ condition.  For a single domain under open boundary condition(OBC), the GBZ condition is $\mu_{\alpha,s_\alpha}=\mu_{\alpha,s_\alpha+1}$~\cite{suppYokomizoNon}, while equalities at other adjacent-root pairs define the auxiliary generalized Brillouin
zone(aGBZ)~\cite{suppYangNon}.  The additional maximal-sector condition \eqref{eq:supp_maximal_w_sector} selects the part of these individual domain GBZ/aGBZ spectra that survives the global DW-ring boundary conditions.

In Case II, no equal-modulus pair is required within any individual domain.  When going from sector $w$ to sector $w-1$, one additional root, $\beta_{\alpha,s_\alpha+w}$, is selected in every domain.  $\lambda_w(E)=0$ means that the exponential growth accumulated by one dominant non-Bloch component in each domain vanishes after one round trip around the ring.  The resulting state is a traveling-wave skin mode selected by the closed-ring geometry and has no counterpart under OBC.

\subsection{An illustrative example}
To illustrate the expansion of the determinant and reduction to the GBZ condition, let us consider three domains with different numbers of roots.  We take
\begin{equation}
    (s_1,p_1)=(1,1),\qquad
    (s_2,p_2)=(1,2),\qquad
    (s_3,p_3)=(2,2).
\end{equation}
where $d_\alpha=s_\alpha+p_\alpha$. Correspondingly, the full boundary-equation matrix is a $9\times9$ matrix. Explicitly, it can be written as
\begin{equation}
M(E)
=
\begin{pmatrix}
l_{1,1}^{(1)}\beta_{1,1}^{N_1}
&
l_{1,2}^{(1)}\beta_{1,2}^{N_1}
&
g_{1,1}^{(2)}
&
g_{1,2}^{(2)}
&
g_{1,3}^{(2)}
&
0&0&0&0
\\
l_{2,1}^{(1)}\beta_{1,1}^{N_1}
&
l_{2,2}^{(1)}\beta_{1,2}^{N_1}
&
g_{2,1}^{(2)}
&
g_{2,2}^{(2)}
&
g_{2,3}^{(2)}
&
0&0&0&0
\\
0&0&
l_{1,1}^{(2)}\beta_{2,1}^{N_2}
&
l_{1,2}^{(2)}\beta_{2,2}^{N_2}
&
l_{1,3}^{(2)}\beta_{2,3}^{N_2}
&
g_{1,1}^{(3)}
&
g_{1,2}^{(3)}
&
g_{1,3}^{(3)}
&
g_{1,4}^{(3)}
\\
0&0&
l_{2,1}^{(2)}\beta_{2,1}^{N_2}
&
l_{2,2}^{(2)}\beta_{2,2}^{N_2}
&
l_{2,3}^{(2)}\beta_{2,3}^{N_2}
&
g_{2,1}^{(3)}
&
g_{2,2}^{(3)}
&
g_{2,3}^{(3)}
&
g_{2,4}^{(3)}
\\
0&0&
l_{3,1}^{(2)}\beta_{2,1}^{N_2}
&
l_{3,2}^{(2)}\beta_{2,2}^{N_2}
&
l_{3,3}^{(2)}\beta_{2,3}^{N_2}
&
g_{3,1}^{(3)}
&
g_{3,2}^{(3)}
&
g_{3,3}^{(3)}
&
g_{3,4}^{(3)}
\\
0&0&
l_{4,1}^{(2)}\beta_{2,1}^{N_2}
&
l_{4,2}^{(2)}\beta_{2,2}^{N_2}
&
l_{4,3}^{(2)}\beta_{2,3}^{N_2}
&
g_{4,1}^{(3)}
&
g_{4,2}^{(3)}
&
g_{4,3}^{(3)}
&
g_{4,4}^{(3)}
\\
g_{1,1}^{(1)}
&
g_{1,2}^{(1)}
&
0&0&0&
l_{1,1}^{(3)}\beta_{3,1}^{N_3}
&
l_{1,2}^{(3)}\beta_{3,2}^{N_3}
&
l_{1,3}^{(3)}\beta_{3,3}^{N_3}
&
l_{1,4}^{(3)}\beta_{3,4}^{N_3}
\\
g_{2,1}^{(1)}
&
g_{2,2}^{(1)}
&
0&0&0&
l_{2,1}^{(3)}\beta_{3,1}^{N_3}
&
l_{2,2}^{(3)}\beta_{3,2}^{N_3}
&
l_{2,3}^{(3)}\beta_{3,3}^{N_3}
&
l_{2,4}^{(3)}\beta_{3,4}^{N_3}
\\
g_{3,1}^{(1)}
&
g_{3,2}^{(1)}
&
0&0&0&
l_{3,1}^{(3)}\beta_{3,1}^{N_3}
&
l_{3,2}^{(3)}\beta_{3,2}^{N_3}
&
l_{3,3}^{(3)}\beta_{3,3}^{N_3}
&
l_{3,4}^{(3)}\beta_{3,4}^{N_3}
\end{pmatrix}.
\label{eq:three_domain_234_matrix}
\end{equation}
Here $l_{i,j}^{(\alpha)}$ and $g_{i,j}^{(\alpha)}$ are matrix elements of $L_\alpha$ and $G_\alpha$, respectively.  The important point is that only the blocks $L_\alpha D_\alpha$ contain the length-dependent factors $\beta_{\alpha,m}^{N_\alpha}$, while the matrices $G_\alpha$ contain no such exponential dependence.

We now expand $\det M(E)$ explicitly. Here $(d_1,d_2,d_3)=(2,3,4)$ and the allowed range of the common winding number is $w_-=-\min_\alpha s_\alpha=-1$, $w_+=\min_\alpha p_\alpha=1$. Therefore, this example contains precisely the three sectors $w=-1,0,1$. In each sector, the number of columns selected from $L_\alpha D_\alpha$ is $k_\alpha=p_\alpha-w$. Consequently, the allowed column selections are
\begin{equation}
\begin{array}{c|c|c}
    w
    &
    (k_1,k_2,k_3)
    &
    (d_1-k_1,d_2-k_2,d_3-k_3)
    \\
    \hline
    -1 & (2,3,3) & (0,0,1)
    \\
     0 & (1,2,2) & (1,1,2)
    \\
     1 & (0,1,1) & (2,2,3)
\end{array}
\label{eq:three_domain_selection_table}
\end{equation}
The second column gives the numbers of columns selected from $L_\alpha D_\alpha$, while the last column gives the corresponding numbers of columns selected from $G_\alpha$.

The determinant can therefore be decomposed into the three $w$ sectors as
\begin{equation}
    \det M(E)=\mathcal D_{-1}(E)+\mathcal D_0(E)+\mathcal D_1(E).
    \label{eq:three_domain_sector_decomposition}
\end{equation}

For $w=-1$, $(k_1,k_2,k_3)=(2,3,3)$, all roots in domains 1 and 2 are selected, while three of the four
roots in domain 3 are selected. The corresponding contribution is
\begin{align}
    \mathcal D_{-1}(E)
    ={}&
    \sum_{1\le i<j<k\le4}
    A_{-1;\{1,2\},\{1,2,3\},\{i,j,k\}}(E)
    \left(
        \beta_{1,1}\beta_{1,2}
    \right)^{N_1}
    \left(
        \beta_{2,1}\beta_{2,2}\beta_{2,3}
    \right)^{N_2}
    \left(
        \beta_{3,i}\beta_{3,j}\beta_{3,k}
    \right)^{N_3}.
    \label{eq:three_domain_w_minus_one}
\end{align}

For $w=0$, we have $(k_1,k_2,k_3)=(1,2,2)$. One root is selected in domain 1, while two roots are selected in each
of domains 2 and 3. Hence
\begin{align}
    \mathcal D_0(E)
    ={}&
    \sum_{i=1}^{2}
    \sum_{1\le j<k\le3}
    \sum_{1\le \ell<m\le4}
    A_{0;\{i\},\{j,k\},\{\ell,m\}}(E)
    \beta_{1,i}^{N_1}
    \left(
        \beta_{2,j}\beta_{2,k}
    \right)^{N_2}
    \left(
        \beta_{3,\ell}\beta_{3,m}
    \right)^{N_3}.
    \label{eq:three_domain_w_zero}
\end{align}

Finally, for $w=1$, we have $(k_1,k_2,k_3)=(0,1,1)$. No root is selected in domain 1, while one root is selected in each of
domains 2 and 3. Therefore,
\begin{align}
    \mathcal D_1(E)
    ={}&
    \sum_{j=1}^{3}
    \sum_{\ell=1}^{4}
    A_{1;\varnothing,\{j\},\{\ell\}}(E)
    \beta_{2,j}^{N_2}
    \beta_{3,\ell}^{N_3}.
    \label{eq:three_domain_w_one}
\end{align}

Eqs.~(\ref{eq:three_domain_w_minus_one})--(\ref{eq:three_domain_w_one}) exhaust all allowed column selections and therefore give the complete length-dependent structure of $\det M(E)$ for this example. The coefficients $A_{w;I_1,I_2,I_3}(E)$ collect all factors that do not scale exponentially with the domain lengths. Individual coefficients may vanish for special choices of the interface matrices.

We now specialize the thermodynamic GBZ condition to this example. Since the GBZ condition depends only on the moduli of the roots, we define $\mu_{\alpha,m}(E)=\log\left|\beta_{\alpha,m}(E)\right|$ with the roots ordered as
\begin{align}
    &\mu_{1,1}\le\mu_{1,2},\notag\\
    &\mu_{2,1}\le\mu_{2,2}\le\mu_{2,3},\notag\\
    &\mu_{3,1}\le\mu_{3,2}\le\mu_{3,3}\le\mu_{3,4}.
\end{align}

For this three-domain example, the set $\zeta$ can be decomposed into
the three common-winding sectors
\begin{equation}
    \zeta
    =
    \zeta_{-1}\cup\zeta_0\cup\zeta_1.
    \label{eq:three_domain_zeta_decomposition}
\end{equation}

For $w=-1$, all roots in domains 1 and 2 are selected, while three roots
are selected in domain 3. Therefore,
\begin{align}
    \zeta_{-1}
    =
    \Bigg\{&
    r_1\left(\mu_{1,1}+\mu_{1,2}\right)
    +
    r_2\left(\mu_{2,1}+\mu_{2,2}+\mu_{2,3}\right)
    \nonumber\\
    &+
    r_3\left(\mu_{3,i}+\mu_{3,j}+\mu_{3,k}\right)
    \,\Bigg|\,
    1\le i<j<k\le4
    \Bigg\}.
    \label{eq:three_domain_zeta_minus_one}
\end{align}

For $w=0$, one root is selected in domain 1, while two roots are
selected in each of domains 2 and 3. Hence,
\begin{align}
    \zeta_0
    =
    \Bigg\{&
    r_1\mu_{1,i}
    +
    r_2\left(\mu_{2,j}+\mu_{2,k}\right)
    +
    r_3\left(\mu_{3,\ell}+\mu_{3,m}\right)
    \,\Bigg|\,
    \nonumber\\
    &1\le i\le2,\qquad
    1\le j<k\le3,\qquad
    1\le\ell<m\le4
    \Bigg\}.
    \label{eq:three_domain_zeta_zero}
\end{align}

For $w=1$, no root is selected in domain 1, while one root is selected
in each of domains 2 and 3. Thus,
\begin{equation}
    \zeta_1
    =
    \left\{
    r_2\mu_{2,j}
    +
    r_3\mu_{3,\ell}
    \,\middle|\,
    1\le j\le3,\quad
    1\le\ell\le4
    \right\}.
    \label{eq:three_domain_zeta_one}
\end{equation}

Because the roots are ordered increasingly by modulus, the largest
element in each sector is obtained by selecting the roots with the
largest values of $\mu_{\alpha,m}$. The three sector maxima are
therefore
\begin{align}
    \Lambda_{-1}
    ={}&
    r_1\left(\mu_{1,1}+\mu_{1,2}\right)
    +
    r_2\left(\mu_{2,1}+\mu_{2,2}+\mu_{2,3}\right)
    \nonumber\\
    &+
    r_3\left(\mu_{3,2}+\mu_{3,3}+\mu_{3,4}\right),
    \label{eq:three_domain_Lambda_minus_one}
    \\
    \Lambda_0
    ={}&
    r_1\mu_{1,2}
    +
    r_2\left(\mu_{2,2}+\mu_{2,3}\right)
    +
    r_3\left(\mu_{3,3}+\mu_{3,4}\right),
    \label{eq:three_domain_Lambda_zero}
    \\
    \Lambda_1
    ={}&
    r_2\mu_{2,3}
    +
    r_3\mu_{3,4}.
    \label{eq:three_domain_Lambda_one}
\end{align}

The differences between adjacent sector maxima are
\begin{align}
    \lambda_0
    &=
    \Lambda_{-1}-\Lambda_0
    =
    r_1\mu_{1,1}
    +
    r_2\mu_{2,1}
    +
    r_3\mu_{3,2},
    \label{eq:three_domain_lambda_zero}
    \\
    \lambda_1
    &=
    \Lambda_0-\Lambda_1
    =
    r_1\mu_{1,2}
    +
    r_2\mu_{2,2}
    +
    r_3\mu_{3,3}.
    \label{eq:three_domain_lambda_one}
\end{align}
The root ordering ensures that
\begin{align}
    \lambda_1-\lambda_0
    ={}&
    r_1\left(\mu_{1,2}-\mu_{1,1}\right)
    +
    r_2\left(\mu_{2,2}-\mu_{2,1}\right)
    \nonumber\\
    &+
    r_3\left(\mu_{3,3}-\mu_{3,2}\right)
    \ge0.
\end{align}
Therefore, the condition that a given sector be a maximal sector can
be written explicitly as
\begin{align}
    w=-1:\qquad&
    \lambda_0\ge0,
    \label{eq:three_domain_maximal_minus_one}
    \\
    w=0:\qquad&
    \lambda_0\le0\le\lambda_1,
    \label{eq:three_domain_maximal_zero}
    \\
    w=1:\qquad&
    \lambda_1\le0.
    \label{eq:three_domain_maximal_one}
\end{align}

We can now write the GBZ condition explicitly for this example. First,
consider an internal degeneracy within a maximal common-winding sector,
corresponding to Case I.

For $w=-1$, all roots in domains 1 and 2 are selected, so an internal
degeneracy can occur only at the selection boundary in domain 3. The
condition is
\begin{equation}
    \lambda_0(E)\ge0,
    \qquad
    \mu_{3,1}(E)=\mu_{3,2}(E).
    \label{eq:three_domain_caseI_minus_one}
\end{equation}

For $w=0$, the selection boundaries occur between roots 1 and 2 in
domains 1 and 2, and between roots 2 and 3 in domain 3. Hence,
\begin{align}
    &\lambda_0(E)\le0\le\lambda_1(E),
    \nonumber\\
    &\mu_{1,1}(E)=\mu_{1,2}(E)
    \quad\text{or}\quad
    \mu_{2,1}(E)=\mu_{2,2}(E)
    \quad\text{or}\quad
    \mu_{3,2}(E)=\mu_{3,3}(E).
    \label{eq:three_domain_caseI_zero}
\end{align}

For $w=1$, no root is selected in domain 1. The selection boundaries
occur between roots 2 and 3 in domain 2 and between roots 3 and 4 in
domain 3. Therefore,
\begin{align}
    &\lambda_1(E)\le0,
    \nonumber\\
    &\mu_{2,2}(E)=\mu_{2,3}(E)
    \quad\text{or}\quad
    \mu_{3,3}(E)=\mu_{3,4}(E).
    \label{eq:three_domain_caseI_one}
\end{align}

Second, Case II corresponds to a degeneracy between two adjacent
common-winding sectors. The degeneracy between the $w=-1$ and $w=0$
sectors is
\begin{equation}
    \lambda_0(E)
    =
    r_1\mu_{1,1}(E)
    +
    r_2\mu_{2,1}(E)
    +
    r_3\mu_{3,2}(E)
    =
    0.
    \label{eq:three_domain_caseII_zero}
\end{equation}
Similarly, the degeneracy between the $w=0$ and $w=1$ sectors is
\begin{equation}
    \lambda_1(E)
    =
    r_1\mu_{1,2}(E)
    +
    r_2\mu_{2,2}(E)
    +
    r_3\mu_{3,3}(E)
    =
    0.
    \label{eq:three_domain_caseII_one}
\end{equation}

Combining the two possibilities, the GBZ condition for this example is
the union of
\begin{align}
    \mathrm{Case\ I:}\qquad
    &\begin{cases}
    \lambda_0\ge0,
    \quad
    \mu_{3,1}=\mu_{3,2},
    &w=-1,
    \\[2mm]
    \lambda_0\le0\le\lambda_1,
    \quad
    \mu_{1,1}=\mu_{1,2}
    \ \text{or}\
    \mu_{2,1}=\mu_{2,2}
    \ \text{or}\
    \mu_{3,2}=\mu_{3,3},
    &w=0,
    \\[2mm]
    \lambda_1\le0,
    \quad
    \mu_{2,2}=\mu_{2,3}
    \ \text{or}\
    \mu_{3,3}=\mu_{3,4},
    &w=1,
    \end{cases}
    \label{eq:three_domain_caseI_complete}
    \\
    \mathrm{Case\ II:}\qquad
    &\lambda_0(E)=0
    \quad\text{or}\quad
    \lambda_1(E)=0.
    \label{eq:three_domain_caseII_complete}
\end{align}
Case I describes standing-wave branches generated by an equal-modulus
pair within one of the three domains, subject to the condition that the
corresponding common-winding sector is maximal. Case II describes
traveling-wave branches arising from equal dominance of two adjacent
common-winding sectors.

\section{SIV.~Domain-Wall-Ring GBZ From Ronkin Formulation}
\label{sec:supp_ronkin}

In this section, we define the constrained Ronkin function for DW ring systems and derive the DW-ring GBZ condition.

\subsection{Derivation of domain-wall-ring GBZ condition}

In this subsection, we first recall that the single-domain Ronkin function is piecewise linear in $\mu$ space, with its slope given by the point-gap winding number. We then impose the global imaginary-gauge constraint of the DW ring and show that a flat region of the constrained Ronkin function is formed when all domains lie in the same winding sector. By rewriting this flat region as a set of simultaneous inequalities and determining when its boundaries become saturated, we derive the conditions under which the flat region loses its finite interior, leading to the standing-wave and traveling-wave branches of the DW-ring GBZ. 

We define the Ronkin function for domain $\alpha$ as~\cite{suppWangAmoeba}
\begin{equation}
    R_\alpha(\mu_\alpha;E)
    =
    \int_0^{2\pi}
    \frac{dk_\alpha}{2\pi}
    \log
    \left|
   f_\alpha(\beta_\alpha;E)
    \right|,
\end{equation}
where $f_\alpha(\beta_\alpha;E) = \det [h_\alpha(e^{ik_\alpha+\mu_\alpha})-E]$. We then define the Ronkin function for DW ring as
\begin{equation}
    \mathcal R(\bm\mu;E)
    =
    \sum_{\alpha=1}^{n-1}r_\alpha R_\alpha(\mu_\alpha;E)
    +
    r_n
    R_n\left(
    -\sum_{\alpha=1}^{n-1}
    \frac{r_\alpha}{r_n}\mu_\alpha;E
    \right),
    \label{eq:supp_CRonkin}
\end{equation}
Let $\beta_{\alpha,m}(E)$ be the roots of the characteristic equation $f_\alpha(\beta_\alpha;E)=\sum_{n=-s_\alpha}^{p_\alpha}a_{\alpha,n}(E)\beta^n_\alpha=0$. Sorting the roots $\beta_{\alpha,m}(E)$ of the characteristic equation $f_\alpha(\beta_\alpha;E)=0$ by their modulus, we have
\begin{align}
\mu_{\alpha,1}(E)\le \mu_{\alpha,2}(E)\le \cdots\le\mu_{\alpha,s_\alpha+p_\alpha}(E),
\end{align}
where $\mu_{\alpha,m}=\log|\beta_{\alpha,m}|$. For simplicity, in the following, we denote $\mu_{\alpha,0}=-\infty$ and $\mu_{\alpha,s_\alpha+p_\alpha+1}=+\infty$.

We first recall the relation between the single-domain Ronkin function and the point-gap winding. For fixed $E$, and away from $\{\mu_{\alpha,m}\}$, the derivative of $R_\alpha$ with respect to $\mu_\alpha$ equals the point-gap winding of the deformed Bloch Hamiltonian $h_\alpha(\beta_\alpha)$ around $E$, as defined in Eq.~(\ref{eq:supp_winding}),
\begin{equation}
    \partial_{\mu_\alpha}R_\alpha(\mu_\alpha;E)
    =
    \frac{1}{2\pi}\int_0^{2\pi}\mathrm{d}k_\alpha\,
    \partial_{\mu_\alpha}\log|f_\alpha(\beta_\alpha;E)|
    =
    \frac{1}{2\pi i}\int_0^{2\pi}\mathrm{d}k_\alpha\,
    \partial_{k_\alpha}\log f_\alpha(\beta_\alpha;E)
    = w_\alpha(\mu_\alpha;E),
\end{equation}
Also, applying Jensen's formula to the single-domain Ronkin function~\cite{suppWangAmoeba}, one obtains
\begin{equation}
    R_\alpha(\mu_\alpha;E)
    =
    \log|a_{\alpha,-s_\alpha}|
    -s_\alpha\mu_\alpha
    +\sum_{n=1}^{l}(\mu_\alpha-\mu_{\alpha,n}),
    \qquad
    \mu_{\alpha,l}\le \mu_\alpha\le \mu_{\alpha,l+1}.
\end{equation}
Here $a_{\alpha,-s_\alpha}$ is the coefficient of $\beta_\alpha^{-s_\alpha}$ in $f_\alpha(\beta_\alpha;E)$. Thus $R_\alpha$ is piecewise linear in $\mu_\alpha$, with slope
\begin{equation}
    \partial_{\mu_\alpha}R_\alpha(\mu_\alpha;E)
    =
    -s_\alpha+l
    =
    w_\alpha(\mu_\alpha;E).
\end{equation}
Equivalently, the interval in which the point-gap winding of domain $\alpha$ takes the value $w_\alpha=-s_\alpha,\ldots,p_\alpha$ is
\begin{equation}
    \mu_\alpha\in
    (\mu_{\alpha,s_\alpha+w_\alpha},\mu_{\alpha,s_\alpha+w_\alpha+1}).
\end{equation}

Taking the derivative of the constrained Ronkin function Eq.~(\ref{eq:supp_CRonkin}) with respect to an independent variable $\mu_\alpha$ gives
\begin{equation}
    \partial_{\mu_\alpha}\mathcal R
    =
    r_\alpha
    \left[
    w_\alpha(\mu_\alpha;E)
    -
    w_n(\mu_n;E)
    \right].
\end{equation}
Thus the flat region of the constrained Ronkin function is the set of imaginary gauge transformations that remove all winding mismatches, namely for all $\alpha,\,w_\alpha(\mu_\alpha;E)=w$. $w$ is chosen from the intersection of the winding-number ranges of all domains, which is
\begin{equation}
     w=-s_*,-s_*+1,\ldots,p_*,
\end{equation}
where $ s_*=\min_\alpha s_\alpha, p_*=\min_\alpha p_\alpha $. For a given $w$, we define the domain-dependent root index $m_\alpha(w)=s_\alpha+w$. Then, when the $\alpha$th domain lies in the sector with winding number $w$,
the corresponding allowed interval of $\mu_\alpha$ is $\left[\mu_{\alpha,m_\alpha(w)},\mu_{\alpha,m_\alpha(w)+1}\right].$ For a given reference energy $E$, flat region of $\mathcal R(\boldsymbol\mu;E)$ occurs when all domains share the same winding number $w$, namely $w_1(\mu_1;E)=w_2(\mu_2;E)=\cdots=w_n(\mu_n;E)=w$, which gives $n$ inequalities on $\boldsymbol{\mu}$,
\begin{equation}
 \exists w,\,   \begin{cases}
        \mu_{\alpha,m_\alpha(w)}\le \mu_\alpha\le\mu_{\alpha,m_\alpha(w)+1} ,\quad \alpha\neq n\\
        \mu_{n,m_n(w)}\le \mu_n=-\sum_{\alpha=1}^{n-1}\frac{r_\alpha}{r_n}\mu_{\alpha}\le\mu_{n,m_n(w)+1}.
    \end{cases}
    \label{eq:flat_region}
\end{equation}
Energies outside the DW ring spectrum correspond to flat regions with finite interior, for which the decay rates are not selected. Spectral energies arise when this flat region collapses, selecting the allowed piecewise imaginary momenta.

We now show how the collapse of this flat region yields the DW ring GBZ conditions.  In particular, we will see that a collapse to an isolated point corresponds to a traveling-wave solution, whereas a generic dimensional reduction of the flat region corresponds to a standing-wave solution.

To analyze the collapse of the flat region, we project Eq.~(\ref{eq:flat_region}) onto each independent variable $\mu_\alpha$, which gives
\begin{align}
    \exists w,\quad \text{such that for all } \alpha\neq n   \, \begin{cases}
        \mu_{\alpha,m_\alpha(w)}\le \mu_\alpha\le\mu_{\alpha,m_\alpha(w)+1}\\
      \mu_{\alpha,m_\alpha(w)+1}-\sum_{\gamma=1}^n\frac{r_\gamma}{r_\alpha}\mu_{\gamma,m_\gamma(w)+1}\le\mu_\alpha\le \mu_{\alpha,m_\alpha(w)}-\sum_{\gamma=1}^n\frac{r_\gamma}{r_\alpha}\mu_{\gamma,m_\gamma(w)}.
    \end{cases}
    \label{eq:flat_region2}
\end{align}
The condition for the system of inequalities to admit a solution is
\begin{equation}
  \text{for all }\alpha,\quad  \begin{cases}
        \mu_{\alpha,m_\alpha(w)}\le  \mu_{\alpha,m_\alpha(w)}-\sum_{\gamma=1}^n\frac{r_\gamma}{r_\alpha}\mu_{\gamma,m_\gamma(w)},\\
        \mu_{\alpha,m_\alpha(w)+1}-\sum_{\gamma=1}^n\frac{r_\gamma}{r_\alpha}\mu_{\gamma,m_\gamma(w)+1}\le \mu_{\alpha,m_\alpha(w)+1},
    \end{cases}
\end{equation}
which gives
\begin{equation}
    \sum_{\alpha=1}^n r_\alpha\mu_{\alpha,m_\alpha(w)}\le0\le \sum_{\alpha=1}^n r_\alpha\mu_{\alpha,m_\alpha(w)+1}.
\end{equation}
For each $w$, the collapse of the flat region falls into the following four cases:
\begin{align}
    \text{case 1.}\quad \exists\alpha, \text{such that}\,\,& \begin{cases}
\mu_{\alpha,m_\alpha(w)+1}=\mu_{\alpha,m_\alpha(w)} ,\\
        \mu_{\alpha,m_\alpha(w)+1}\le \mu_{\alpha,m_\alpha(w)}-\sum_{\gamma=1}^n\frac{r_\gamma}{r_\alpha}\mu_{\gamma,m_\gamma(w)},\\
\mu_{\alpha,m_\alpha(w)+1}\ge\mu_{\alpha,m_\alpha(w)+1}-\sum_{\gamma=1}^n\frac{r_\gamma}{r_\alpha}\mu_{\gamma,m_\gamma(w)+1};
    \end{cases}\\
        \text{case 2.}\quad \exists\alpha, \text{such that}\,\, & \begin{cases}
        \mu_{\alpha,m_\alpha(w)+1}=\mu_{\alpha,m_\alpha(w)+1}-\sum_{\gamma=1}^n\frac{r_\gamma}{r_\alpha}\mu_{\gamma,m_\gamma(w)+1},\\
        \mu_{\alpha,m_\alpha(w)+1}\le \mu_{\alpha,m_\alpha(w)}-\sum_{\gamma=1}^n\frac{r_\gamma}{r_\alpha}\mu_{\gamma,m_\gamma(w)},\\
\mu_{\alpha,m_\alpha(w)}\le\mu_{\alpha,m_\alpha(w)+1}-\sum_{\gamma=1}^n\frac{r_\gamma}{r_\alpha}\mu_{\gamma,m_\gamma(w)+1};
    \end{cases}\\
      \text{case 3.}\quad \exists\alpha, \text{such that}\,\, & \begin{cases}
\mu_{\alpha,m_\alpha(w)}=\mu_{\alpha,m_\alpha(w)}-\sum_{\gamma=1}^n\frac{r_\gamma}{r_\alpha}\mu_{\gamma,m_\gamma(w)} ,\\
        \mu_{\alpha,m_\alpha(w)+1}\ge \mu_{\alpha,m_\alpha(w)}-\sum_{\gamma=1}^n\frac{r_\gamma}{r_\alpha}\mu_{\gamma,m_\gamma(w)},\\
        \mu_{\alpha,m_\alpha(w)}\ge\mu_{\alpha,m_\alpha(w)+1}-\sum_{\gamma=1}^n\frac{r_\gamma}{r_\alpha}\mu_{\gamma,m_\gamma(w)+1};
    \end{cases}\\
    \text{case 4.}\quad \exists\alpha, \text{such that}\,\,& \begin{cases}
        \mu_{\alpha,m_\alpha(w)+1}-\mu_{\alpha,m_\alpha(w)}=\sum_{\gamma=1}^n\frac{r_\gamma}{r_\alpha}(\mu_{\gamma,m_\gamma(w)+1}-\mu_{\gamma,m_\gamma(w)}), \\
        \mu_{\alpha,m_\alpha(w)+1}\ge \mu_{\alpha,m_\alpha(w)}-\sum_{\gamma=1}^n\frac{r_\gamma}{r_\alpha}\mu_{\gamma,m_\gamma(w)},\\
        \mu_{\alpha,m_\alpha(w)}\le\mu_{\alpha,m_\alpha(w)+1}-\sum_{\gamma=1}^n\frac{r_\gamma}{r_\alpha}\mu_{\gamma,m_\gamma(w)+1}.
    \end{cases}
\end{align}
Each case can be further simplified as follows
\begin{align}
  \text{case 1.}\quad \exists\alpha, \text{such that}\,\,& \begin{cases}
\mu_{\alpha,m_\alpha(w)+1}=\mu_{\alpha,m_\alpha(w)} ,\\
\sum_{\gamma=1}^nr_\gamma\mu_{\gamma,m_\gamma(w)}\le0\le\sum_{\gamma=1}^nr_\gamma\mu_{\gamma,m_\gamma(w)+1};
    \end{cases}\\
        \text{case 2.}\quad \exists\alpha, \text{such that}\,\, & \begin{cases}
\sum_{\gamma=1}^nr_\gamma\mu_{\gamma,m_\gamma(w)+1}=0,\\
\mu_{\alpha,m_\alpha(w)}\le\mu_{\alpha,m_\alpha(w)+1}\le\mu_{\alpha,m_\alpha(w)}+\sum_{\gamma=1}^n\frac{r_\gamma}{r_\alpha}(\mu_{\gamma,m_\gamma(w)+1}-\mu_{\gamma,m_\gamma(w)});
    \end{cases}\\
      \text{case 3.}\quad \exists\alpha, \text{such that}\,\, & \begin{cases}
\sum_{\gamma=1}^nr_\gamma\mu_{\gamma,m_\gamma(w)}=0,\\
\mu_{\alpha,m_\alpha(w)}\le\mu_{\alpha,m_\alpha(w)+1}\le\mu_{\alpha,m_\alpha(w)}+\sum_{\gamma=1}^n\frac{r_\gamma}{r_\alpha}(\mu_{\gamma,m_\gamma(w)+1}-\mu_{\gamma,m_\gamma(w)});
    \end{cases}\\
    \text{case 4.}\quad \exists\alpha, \text{such that}\,\,& \begin{cases}\forall \gamma\neq \alpha,\quad 
\mu_{\gamma,m_\gamma(w)+1}=\mu_{\gamma,m_\gamma(w)} ,\\
\sum_{\gamma=1}^nr_\gamma\mu_{\gamma,m_\gamma(w)}\le0\le\sum_{\gamma=1}^nr_\gamma\mu_{\gamma,m_\gamma(w)+1}.
    \end{cases}
\end{align}
Since $\mu_{\alpha,m}$ are ordered by magnitude, $\mu_{\gamma,m_\gamma(w)+1}\ge\mu_{\gamma,m_\gamma(w)}$, the second inequality in the reduced forms of cases 2 and 3 holds automatically.

Further, if case 2 or 3 is satisfied for a given $w$, Eq.~(\ref{eq:flat_region2}) becomes
\begin{align}
 \text{for all }\alpha\neq n   \,& \begin{cases}
        \mu_{\alpha,m_\alpha(w)}\le \mu_\alpha\le\mu_{\alpha,m_\alpha(w)+1}\\
      \mu_{\alpha,m_\alpha(w)+1}\le\mu_\alpha\le \mu_{\alpha,m_\alpha(w)}-\sum_{\gamma=1}^n\frac{r_\gamma}{r_\alpha}\mu_{\gamma,m_\gamma(w)};
    \end{cases}\\
    \text{or}\quad & \begin{cases}
        \mu_{\alpha,m_\alpha(w)}\le \mu_\alpha\le\mu_{\alpha,m_\alpha(w)+1}\\
      \mu_{\alpha,m_\alpha(w)+1}-\sum_{\gamma=1}^n\frac{r_\gamma}{r_\alpha}\mu_{\gamma,m_\gamma(w)+1}\le\mu_\alpha\le \mu_{\alpha,m_\alpha(w)},
    \end{cases}
\end{align}
which means that the flat region collapses to an endpoint of the allowed region. By contrast, in cases 1 and 4, the corresponding flat region is merely reduced in dimension and generically does not collapse to a point.

Combining the four cases above, we arrive at the GBZ condition for the DW ring, Eqs.~(\ref{eq:supp_GBZ_case1})-(\ref{eq:supp_GBZ_case2}). Based on their forms, the solutions are classified into cases I and II, corresponding respectively to standing-wave and traveling-wave solutions.
\begin{align}
  &  \text{Case \rm I.} \quad \,\,\exists w,\,\text{such that}\notag \\
    &\quad \quad\qquad \begin{cases}
        \sum_{\alpha}r_\alpha \mu_{\alpha,m_\alpha(w)}\le 0 \le \sum_{\alpha}r_\alpha \mu_{\alpha,m_\alpha(w)+1},\\
        \exists \alpha,\,\, \mu_{\alpha,m_\alpha(w)} = \mu_{\alpha,m_{\alpha}(w)+1}.
    \end{cases}
    \label{Seq:GBZeq1}\\
    &\text{Case \rm II.} \quad\exists w,\,\text{such that}\,\, \sum_{\alpha}r_\alpha \mu_{\alpha,m_\alpha(w)}=0 ,\label{Seq:GBZeq2}
\end{align}

\subsection{Spectral-potential interpretation of the constrained Ronkin function}

In this subsection, we interpret the minimum of the constrained Ronkin function as the thermodynamic spectral potential of the DW ring and numerically verify this interpretation by comparing the resulting density of states with that obtained from real-space diagonalization.
We first recall the electrostatic interpretation used in the Amoeba formulation of non-Bloch band theory~\cite{suppWangAmoeba}.  For a finite non-Hermitian Hamiltonian with eigenvalues $\{\epsilon_m\}$ and total system size $N$, one may define the spectral Coulomb potential on the complex-energy plane as
\begin{equation}
    \Phi_{\rm diag}^{(N)}(E)
    =
    \frac{1}{N}
    \sum_m
    \log |E-\epsilon_m| .
\end{equation}
In the thermodynamic limit, the density of states(DOS) is obtained from this potential by taking the Laplacian,
\begin{equation}
    \rho(E)
    =
    \frac{1}{2\pi}
    \Delta_E \Phi(E),
\end{equation}
where $\Delta_E=\partial_{\mathrm{Re}E}^2+\partial_{\mathrm{Im}E}^2$.  For translationally invariant non-Hermitian systems, Ref.~\cite{suppWangAmoeba} showed in the Amoeba formulation that the thermodynamic spectral potential can be obtained from the minimum of the Ronkin function, namely
\begin{equation}
    \Phi(E)
    =
    \phi(E),
    \qquad
    \phi(E)
    =
    \min_{\mu} R(\mu;E).
\end{equation}
Thus the Ronkin function also encodes the spectral density through its minimum value.
\begin{figure}
    \centering
    \includegraphics[width=.6\linewidth]{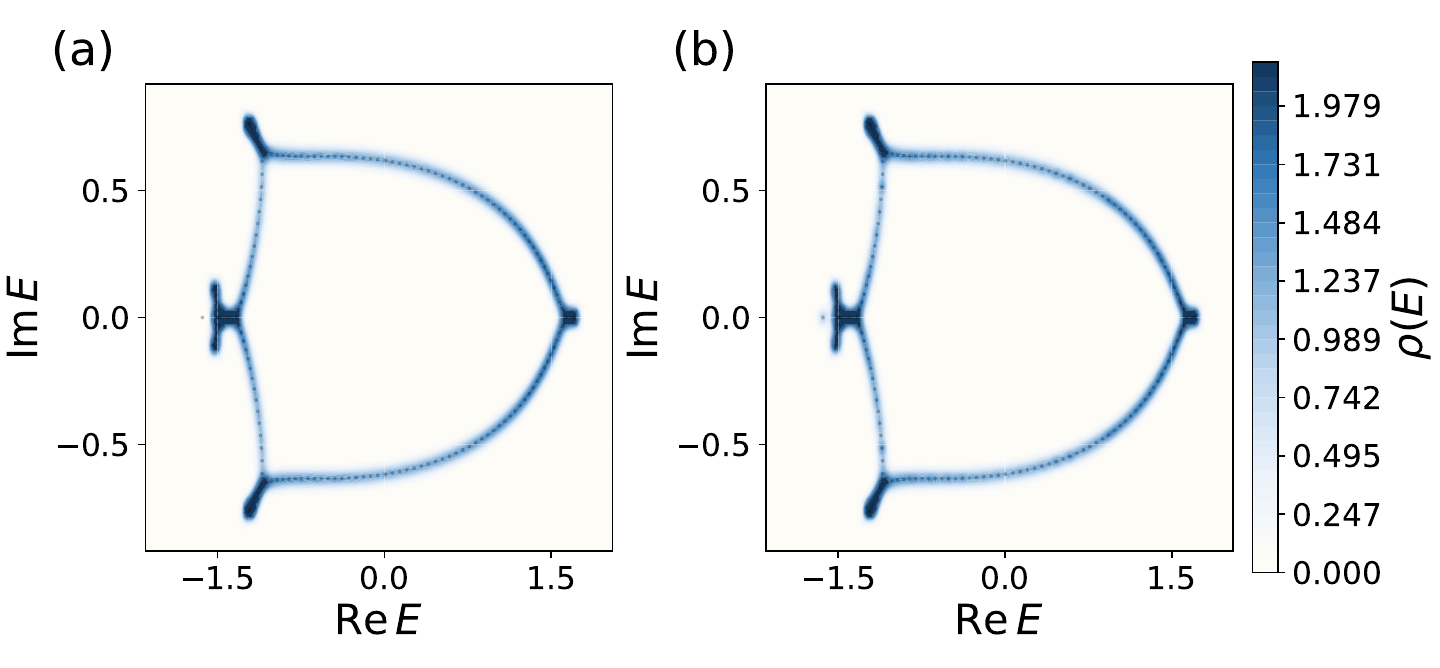}
    \caption{Numerical check of the spectral-potential interpretation of the constrained Ronkin function. (a) Density of states obtained from the constrained Ronkin minimum, (b) Density of states obtained from real-space diagonalization.
The agreement between the two panels supports the interpretation of the constrained Ronkin minimum as the spectral Coulomb potential of the DW ring in the thermodynamic limit. Model and parameters are given in the main text.}
    \label{FigS1}
\end{figure}

As a consistency check of the constrained Ronkin formulation for DW ring, we further numerically examine its spectral-potential interpretation.  Motivated by the Amoeba formulation of translationally invariant systems, we define
\begin{equation}
    \phi_{\rm DW}(E)
    =
    \min_{\boldsymbol{\mu}}
    \mathcal R(\boldsymbol{\mu};E),
\end{equation}
and compare the density obtained from its Laplacian,
\begin{equation}
    \rho_{\rm DW}^{\rm Ronkin}(E)=\frac{1}{2\pi} \Delta_E \phi_{\rm DW}(E),
\end{equation}
with the density extracted from the real-space spectrum,
\begin{equation}
    \rho_{\rm DW}^{\rm diag}(E) =\frac{1}{2\pi}\Delta_E\left[ \frac{1}{N}\sum_m \log |E-\epsilon_m| \right].
\end{equation}
Here $\{\epsilon_m\}$ are the eigenvalues of the finite DW ring Hamiltonian $H_{\rm DW}$, and the Laplacian is evaluated on the complex-energy plane. As shown in Fig.~\ref{FigS1}, the two DOS agree well within the numerical resolution of the energy grid.  This agreement provides a consistency check that the constrained Ronkin minimum plays the role of the spectral Coulomb potential for the DW ring in the thermodynamic limit.

\section{SV.~GBZ for Open Domain-Wall Chains}
\label{sec:supp_open_chain}

For comparison, in this section, we consider an open domain-wall chain(DW chain), where the domains are connected in series but the two ends are left open.

In this case, at both ends, the end-to-end coupling terms of the DW ring are absent, the interface-coupling block matrix $G_1$ and $L_n$ in Eq.~(\ref{eq:supp_M}) are replaced by open-boundary block matrix $B_1$ and $B_n$, respectively. The coefficient matrix then becomes
\begin{equation}
    M_{\rm OBC}(E)
    =
    \begin{pmatrix}
    B_1  & 0& 0&\cdots&0&0\\
      L_1  D_1 &G_2 & 0&\cdots & 0 & 0\\
        0 &L_2 D_2 & G_3 & \cdots & 0 & 0\\
        0 & 0 & L_3 D_3 & \ddots &0 & 0\\
        \vdots & \vdots &\vdots & \ddots & \ddots & \vdots\\
        0 & 0 & 0 &\cdots &L_{n-1}D_{n-1} &G_n \\
  0 & 0  & 0 & \cdots & 0 &B_nD_n
    \end{pmatrix}.
    \label{eq:supp_Mobc}
\end{equation}
Here $B_1$ and $B_nD_n$ are $s_1\times d_1$ and $p_n\times d_n$, respectively.

Following the same method, the determinant takes the form
\begin{equation}
      \det M_{\rm OBC}(E)
      =
      \sum_{\substack{
      I_1,\ldots,I_n\\
      |I_\alpha|=p_\alpha
      }}
      A_{I_1,\ldots,I_n}(E)
      \prod_{\alpha=1}^{n}
      \left(
          \prod_{m\in I_\alpha}
          \beta_{\alpha,m}(E)
      \right)^{N_\alpha}.
      \label{eq:supp_detM_obc_general}
\end{equation}
Here $I_\alpha\subset\{1,2,\ldots,d_\alpha\}$, and the coefficients $A_{I_1,\ldots,I_n}(E)$ contain the boundary details but no length-dependent exponential factors.

Compared with the end-to-end connected DW ring geometry, the open DW-chain fixes the number of roots selected from each domain to be $p_\alpha$.  In the notation of the DW ring determinant expansion Eq.~(\ref{eq:supp_detM_expansion}), this corresponds to selecting only the $w=0$ sector.

We denote the set of length-dependent exponential factors by
\begin{equation}
    \eta_{\rm OBC}
    =
    \left\{
    \prod_{\alpha=1}^{n}
    \left(
        \prod_{m\in I_\alpha}
        \beta_{\alpha,m}(E)
    \right)^{N_\alpha}
    \,\middle|\,
    |I_\alpha|=p_\alpha,\;
    I_\alpha\subset \{1,2,\ldots,d_\alpha\}
    \right\}.
\end{equation}
The GBZ condition for continuous band requires that the maximal modulus in $\eta_{\rm OBC}$ be attained by at least two terms.

Since the roots in each domain are ordered as
\begin{equation}
    \mu_{\alpha,1}
    \le
    \mu_{\alpha,2}
    \le
    \cdots
    \le
    \mu_{\alpha,d_\alpha},
    \qquad
    \mu_{\alpha,m}=\log|\beta_{\alpha,m}|,
\end{equation}
the unique largest term, if nondegenerate, is obtained by selecting the $p_\alpha$ largest roots in every domain, namely
\begin{equation}
    \mu_{\alpha,s_\alpha+1},
    \mu_{\alpha,s_\alpha+2},
    \ldots,
    \mu_{\alpha,d_\alpha}.
\end{equation}
This maximal term becomes degenerate precisely when, for at least one domain $\alpha$, the boundary between the selected and unselected roots is degenerate $\mu_{\alpha,s_\alpha}(E)=\mu_{\alpha,s_\alpha+1}(E)$.
Thus the thermodynamic GBZ condition for the DW-chain is
\begin{equation}
    \exists \alpha,
    \qquad
    \mu_{\alpha,s_\alpha}(E)
    =
    \mu_{\alpha,s_\alpha+1}(E).
    \label{eq:supp_DW_chain_GBZ_general}
\end{equation}

\begin{figure}[H]
    \centering
    \includegraphics[width=0.6\linewidth]{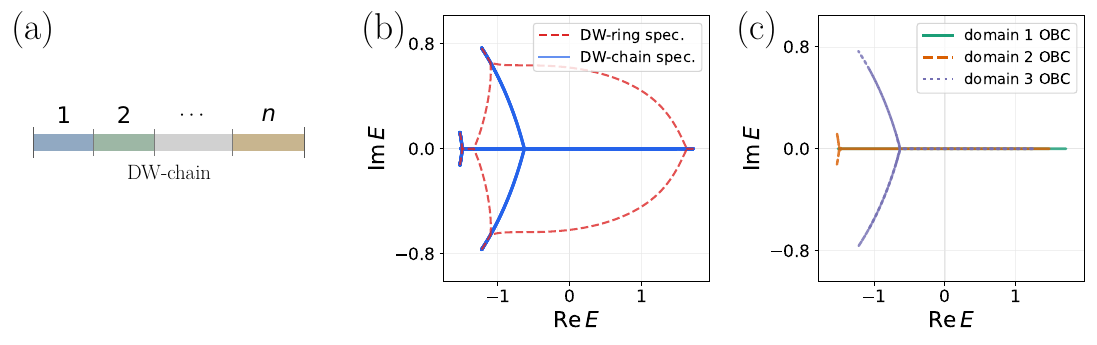}
    \caption{Spectrum of the open DW chain. 
(a) Schematic of a general $n$-domain DW chain obtained by opening the DW ring. 
(b) Comparison between the thermodynamic spectrum of the DW ring and that of the corresponding DW chain obtained by cutting the $3|1$ interface. The red dashed curve denotes the DW-ring spectrum, while the blue curve denotes the DW-chain spectrum. 
(c) Individual-domain OBC spectra of the three constituent domains. Their union gives the DW-chain spectrum in (b). Model and parameters are given in the main text.
}
    \label{figS2}
\end{figure}

Therefore, unlike the DW ring, the DW chain has no adjacent-sector degeneracy associated with a round-trip condition.  Its thermodynamic spectrum is formed only by standing-wave solutions.  The GBZ condition above is the domain-wall generalization of the conventional OBC GBZ condition for a translationally invariant chain. For the model studied in the main text, the spectra of DW chain and DW ring are demonstrated in Fig.~\ref{figS2}.  The DW-chain GBZ condition also shows that a DW-chain supports no traveling-wave sector in the thermodynamic limit, which highlights the sensitivity to the boundary condition in domain-wall systems.

We also provide an alternative method to determine the GBZ condition of a DW-chain by generalizing the Ronkin-function formalism. As before, we introduce a piecewise imaginary gauge transformation.  For a given energy, the goal is to find a set of imaginary gauges such that the transformed system no longer exhibits NHSE.  These parameters then give precisely the decay rates of the corresponding domain-wall eigenstate. 

The absence of boundary topologically protected skin modes after imaginary gauge transformation imposes $w_1(\mu_1;E)\leq0$ at the left end and $w_n(\mu_n;E)\ge0$ at the right end according to the topological origin of NHSE in open-boundary systems~\cite{suppOkumaTopological}. Meanwhile, the domain-wall NHSE is governed by the mismatch of the winding across an interface; hence no skin effect occurs at an interface only when the two adjacent domains have the same winding number. Combining these constraints, a globally NHSE-free DW-chain requires the winding number of the transformed system to be zero in all domains. In this case there is no ring structure and hence no global constraint on the imaginary gauges, to capture the topology of the NHSE in a DW-chain, the corresponding Ronkin function is simply the weighted sum of $R_\alpha,\alpha=1,2,\cdots,n$,
\begin{equation}
    \mathcal R_{\rm OBC}(\tilde {\boldsymbol\mu};E)
    =
    \sum_{\alpha=1}^{n}
    r_\alpha R_\alpha(\mu_\alpha;E),
\end{equation}
where $\tilde{\boldsymbol{\mu}}=(\mu_1,\mu_2,\cdots,\mu_n)$. Since $\partial_{\mu_\alpha} \mathcal{R}_{\rm OBC}=r_\alpha\partial_{\mu_\alpha} R_\alpha=r_\alpha w_\alpha$, the flat region of $\mathcal{R}_{\rm OBC}$ occurs when all the domains are in the zero-winding sector, $w_1(\mu_1;E)=w_2(\mu_2;E)=\cdots=w_n(\mu_n;E)=0$.

Thermodynamic eigenenergy is selected only when this flat region collapses and loses its interior. The flat region collapse condition then reduces to the condition that at least one domain satisfies its own GBZ condition, $ \mu_{\alpha,s_\alpha}(E)=\mu_{\alpha,s_\alpha+1}(E)$. Consequently, in the thermodynamic limit, the bulk spectrum of a DW-chain is given by the union of the open-boundary spectra of its constituent domains.

\section{SVI.~Spectral winding and the traveling-wave sector}

In this section, we explain why the flux spectral winding of a closed DW ring is carried by the traveling-wave sector.  We insert a flux $\Phi$ through the ring and choose a gauge in which the flux appears on a single cut.  The resulting Hamiltonian is denoted by $H_{\rm DW}(\Phi)$.  For a reference energy $E_B$ that remains in a point gap for all $\Phi\in[0,2\pi)$, we define
\begin{equation}
    W_{\rm DW}(E_B)
    =
    \frac{1}{2\pi i}
    \int_0^{2\pi}d\Phi\,
    \partial_\Phi
    \log\det[H_{\rm DW}(\Phi)-E_B].
\end{equation}

Consider a traveling-wave solution in which the $\alpha$th domain is dominated by a single non-Bloch mode
\begin{equation}
    \psi_\alpha(x)
    \sim
    C_\alpha
    \beta_{\alpha,m_\alpha}^{\,x}
    u_{\alpha,m_\alpha},
    \qquad
    \beta_{\alpha,m_\alpha}
    =
    e^{\mu_{\alpha,m_\alpha}+ik_{\alpha,m_\alpha}} .
\end{equation}
Here $u_{\alpha,m_\alpha}$ is the internal eigenvector. When this mode propagates through the $\alpha$th domain, its amplitude acquires the factor $\beta_{\alpha,m_\alpha}^{N_\alpha}$. The local boundary conditions at the domain walls contribute only non-exponential factors.  Denoting their product by $\mathcal A(E)$, the closed-ring condition after one full round trip is
\begin{equation}
    e^{i\Phi}
    \mathcal A(E)
    \prod_{\alpha=1}^{n}
    \beta_{\alpha,m_\alpha}^{N_\alpha}=1 .
\end{equation}
Taking the modulus and dividing by the total length $N=\sum_\alpha N_\alpha$ gives, in the thermodynamic limit,
\begin{equation}
    \sum_{\alpha=1}^{n}
    r_\alpha
    \mu_{\alpha,m_\alpha}(E)
    =
    0 ,
    \qquad
    r_\alpha=\frac{N_\alpha}{N}.
\end{equation}
This is precisely the zero-growth condition of the traveling-wave sector.  Taking the phase gives
\begin{equation}
    \sum_{\alpha=1}^{n}
    N_\alpha k_{\alpha,m_\alpha}(E)
    +
    \arg \mathcal A(E)
    +
    \Phi
    =
    2\pi q,
    \qquad q\in\mathbb Z .
\end{equation}
Hence changing $\Phi$ shifts the corresponding energy $E$.  Therefore a $2\pi$ flux insertion generates a spectral flow along the traveling-wave branch.  If this branch winds around a base point $E_B$, it gives a nonzero contribution to $W_{\rm DW}(E_B)$.

By contrast, a standing-wave branch is selected locally by an equal-modulus pair in at least one domain.  Its existence is controlled by the interference of two non-Bloch components inside that domain and by local boundary conditions at the neighboring domain walls, rather than by propagation around the entire ring.  A global flux only changes the phase accumulated in a full round trip, so it does not affect the standing-wave branch at leading order in the thermodynamic limit. Thus it does not contribute to the flux-induced spectral winding.

Thus the two sectors respond differently to a flux insertion.  Traveling-wave modes are fixed by a global round-trip phase condition and are flux-sensitive; standing-wave modes are fixed by local equal-modulus conditions and do not contribute to the flux spectral winding in the thermodynamic limit.

\bibliographystyle{apsrev4-2}

\end{document}